\documentclass{aa}  

\usepackage{graphicx}
\usepackage{mathtools}
\usepackage[varg]{txfonts}
\usepackage{natbib}
\usepackage{subfig}
\usepackage{bm}
\usepackage{longtable}
\usepackage[colorinlistoftodos]{todonotes}
\usepackage{hyperref}
\usepackage{rotating}
\bibpunct[; ]{(}{)}{,}{a}{}{;}
\newcommand{\bd}{BD+46$^{\circ}$442}
\newcommand{\angstrom}{\AA}

\newcommand{\dd}{\mathrm{d}}
\newcommand{\myr}{M_\odot\,\text{yr}^{-1}}
\newcommand{\rsun}{R$_\odot$\ }
\newcommand{\iras}{IRAS19135+3937}

\begin{document}

   \title{Determining mass accretion and jet mass-loss rates in post-asymptotic giant branch binary systems\thanks{Based on observations made with the Mercator Telescope, operated on the island of La Palma by the Flemish Community, at the Spanish Observatorio del Roque de los Muchachos of the Instituto de Astrofísica de Canarias.}}

   \author{D. Bollen \inst{1,2,3}, D. Kamath \inst{1,2}, O. De Marco \inst{1,2},H. Van Winckel \inst{3}, M. Wardle \inst{1,2}
   }

	\institute{Department of Physics \& Astronomy, Macquarie University, 
              Sydney, NSW 2109, Australia
              \and
              Astronomy, Astrophysics and Astrophotonics Research Centre, 
              Macquarie University, Sydney, NSW 2109, Australia
              \and
              Instituut voor Sterrenkunde (IvS), KU Leuven,
              Celestijnenlaan 200D, B-3001 Leuven, Belgium
               \email{dylan.bollen@kuleuven.be}
              }

   \date{}
   \authorrunning{Bollen et al., }
 
    \abstract
  {}
  {In this study we determine the morphology and mass-loss rate of jets emanating from the companion in post-asymptotic giant branch (post-AGB) binary stars with a circumbinary disk. In doing so, we also determine the mass accretion rates on to the companion and investigate the source feeding the circum-companion accretion disk. 
  }
  {We perform a spatio-kinematic modelling of the jet of two well-sampled post-AGB binaries, \bd\ and \iras, by fitting the orbital phased time series of H$\alpha$ spectra. Once the jet geometry, velocity and scaled density structure are computed, we carry out radiative transfer modelling of the jet for the first four Balmer lines to determine the jet densities, thus allowing us to compute the jet mass-loss rates and mass accretion rates. We distinguish the origin of the accretion by comparing the computed mass accretion rates with theoretically-estimated mass-loss rates, both from the post-AGB star and from the circumbinary disk.}
  {The spatio-kinematic model of the jet reproduces the observed absorption feature in the H$\alpha$ lines. In both objects, the jets have an inner region with extremely low density. The jet model for \bd\, is tilted by $15\degr$ with respect to the orbital axis of the binary system. \iras\, has a smaller tilt of $6\degr$. Using our radiative transfer model, we find the full three-dimensional density structure of both jets. By combining these results, we can compute mass-loss rates of the jets, which are of the order of $10^{-7}-10^{-5}\myr$. From this, we estimate mass accretion rates onto the companion of $10^{-6}-10^{-4}\myr$. 
  }
  {Based on the mass accretion rates found for these two objects, we conclude that the circumbinary disk is most likely the source feeding the circum-companion accretion disk. This is in agreement with the observed depletion patterns in post-AGB binaries, which is caused by re-accretion of gas from the circumbinary disk that is under-abundant in refractory elements. The high accretion rates from the circumbinary disk imply that the lifetime of the disk will be short. Mass-transfer from the post-AGB star cannot be excluded in these systems, but it is unlikely to provide a sufficient mass-transfer rate to sustain the observed jet mass-loss rates.
  }  
  \keywords{Stars: AGB and post-AGB -- 
           Stars: binaries: spectroscopic -- 
           Stars: circumstellar matter --
           Stars: mass-loss --
           ISM: jets and outflows --
           Accretion, accretion disks}
            
   \maketitle
%

\section{Introduction}\label{sec:intro}

Binarity can have a significant impact on the evolution of low-to intermediate mass stars. The binary interactions in these systems can alter their mass loss history, orbital parameters, and lifetimes and can lead to other phenomena such as excretion and accretion disks, jets, and bipolar nebulae \citep{hilditch01}. Post-AGB stars in binary systems are no exception. These are stars of low-to intermediate mass in a final transition phase after the AGB \citep{vanwinckel03}. The luminous post-AGB star in these binary systems is in orbit with a main-sequence (MS) companion of low mass \citep[$0.1 - 2.5\,M_\odot$,][]{oomen18}. Due to their binary interaction history, post-AGB binary systems end up with periods and eccentricities that are currently unexplained by theory \citep{vanwinckel18}.

During the former AGB phase, the star endures a period of mass loss as high as $10^{-4} - 10^{-3}\,\myr$ \citep{ramstedt08}. When in a binary system, the mass loss of the AGB star can be concentrated on the orbital plane of the system, with the bulk of the mass being ejected via the L2 Lagrangian point \citep{hubova19, bermudez20}. The focused mass loss of the star can then become a circumbinary disk \citep{shu79,pejcha16, macleod18}. Observational studies have confirmed the presence of such disks in post-AGB binary systems. Many post-AGB stars have a near-IR dust excess in their spectral energy distribution (SED), which can be explained by dust in the proximity of the central binary system. The observed dust excess is a clear signature of dust residing in a circumbinary disk, close to the system \citep{deruyter06,deroo06,deroo07, kamath14a, kamath15a}. The compact nature of the infrared dust excess has also been confirmed through interferometric studies \citep{bujarrabal13a, hillen13, hillen16a, kluska18}. Additionally, \cite{hillen16a} and \cite{kluska18} identified a flux excess at the location of the companion in the reconstructed interferometric image of post-AGB binary IRAS08544$-$4431. This flux excess is too large to originate from the companion, and most likely stems from an accretion disk around the companion. 

Another commonly observed phenomenon in post-AGB binaries is a high-velocity outflow or jet \citep{gorlova12}. Optical spectra of these objects show a blue-shifted absorption feature in the Balmer lines during superior conjunction, when the companion star is located between the post-AGB star and the observer \citep{gorlova12, gorlova15}, as can be seen in Fig.~\ref{fig:supinf}. The absorption feature in the Balmer lines is interpreted in terms of a jet launched from the vicinity of the companion, that scatters the continuum light from the post-AGB star travelling towards the observer during this phase in the binary orbit \citep{gorlova12}. Due to the orbital motion of the binary, the photospheric light of the post-AGB star shines through various parts of the jet, providing a tomography of the jet. Hence, the orbital-phase dependent variations in the Balmer lines of these jet-creating post-AGB binaries contain an abundance of information about the jet and the binary system \citep{bollen17, bollen19}.

 The jets are likely launched by an accretion disk around the companion. An unknown component in these jet-creating post-AGB binaries is the source feeding the circum-companion accretion disk that launches the jet. Direct observations of the mass-transfer to the circum-companion disk do not exist. The two plausible sources are the post-AGB star, which could transfer mass via the first Lagrangian point (L1) to the companion, or the re-accretion from the circumbinary disk around the system. While this mass transfer has not yet been observed directly, we observe refractory element depletion in the atmosphere of post-AGB stars in binary systems \citep{waters92, vanwinckel95, gezer15, kamath19}. It is suggested that this depletion pattern is caused by re-accretion of circumbinary gas, which is depleted of refractory elements by the formation of dust in the disk. \cite{oomen19} modelled this depletion pattern by implementing the re-accretion of metal-poor gas in their evolutionary models obtained using the Modules for Experiments in Stellar Astrophysics (\texttt{MESA}) code. They compared these models with 58 observed post-AGB stars and found that initial mass accretion rates must be greater than $3\times10^{-7}\myr$ in order to obtain the observed depletion patterns.

 In our previous study \citep{bollen17}, we used the time-series of H$\alpha$ line profiles to show that jets in post-AGB binaries are wide and can reach velocities of $\sim 700\,$km s$^{-1}$. These velocities are of the order of the escape velocity of a MS star, pointing to the nature of the companion. In our recent study \citep{bollen19}, we created a more sophisticated spatio-kinematic model for the jets, from which we determined the jets' geometry, velocity, and scaled density structure. 
 
 In this paper, we will fully exploit the potential of the tomography of the jet from the first four Balmer lines: H$\alpha$, H$\beta$, H$\gamma$, and H$\delta$. We will compute a radiative transfer model of the jet, with the aim of estimating the mass-loss rate of the jet. We do this in two main parts: (1) the spatio-kinematic modelling, as described by \cite{bollen19} and (2) the radiative transfer modelling of the jet. Here, we will focus on part II and the mass ejection and accretion rates. We choose two well-sampled, jet-launching post-AGB binaries for our analysis: \bd\, \citep{gorlova12, bollen17} and \iras\, \citep{gorlova15, bollen19}. Both objects have been observed for the past ten years with the HERMES spectrograph, mounted on the \textit{Mercator} telescope, La Palma, Spain \citep{raskin11}, providing a good amount of data covering the orbital phase of the binary.

The paper is organised as follows:  We describe the methods of our spatio-kinematic modelling and radiative transfer modelling in Sect.~\ref{sec:methods}. We present the results for \bd\, and \iras\, in Sect.~\ref{sec:jetmodelBD} and Sect.~\ref{sec:jetmodelIRAS}, respectively. We discuss these results in Sect.~\ref{sec:discussion} and give a conclusion and summary in Sect.~\ref{sec:conclusions}.

\section{Methods}\label{sec:methods}
 In this study, we expand on the spatio-kinematic model carried out in our previous work \citep{bollen19} by adding new components in the jet structure and we include a new radiative transfer model. Splitting the calculations in two parts, i.e. the spatio-kinematic modelling and the radiative transfer modelling, allows us to fit the jet structure and obtain the jet mass-loss rates. In the following sub-sections, we give a short description of the spatio-kinematic modelling part of the fitting, including improvements of the technique pioneered by \cite{bollen19}, followed by a description of the new radiative transfer modelling.

\subsection{Spatio-kinematic modelling of the jet}\label{ssec:spatio}

To obtain the geometry and kinematics of the jet, we follow the model-fitting routine used by \cite{bollen19}. In brief, we create a spatio-kinematic model of the jet, from which we reproduce the absorption features in the H$\alpha$ line. The modelled lines are then fitted to the observations. To fit our model to the data, we use the \texttt{emcee}-package, which applies an MCMC algorithm \citep{foreman13}. This gives us the best-fitting parameters for the jet.

The model consists of three main components: the post-AGB star, the MS companion, and the jet. The location of the post-AGB star and the companion are determined for each orbital phase by the orbital parameters listed in Table~\ref{tab:orbpar}. The jet in the model is a double cone, centred on the companion. The post-AGB star is approximated as a uniform, flat disk facing the observer. We trace the light travelling from the post-AGB star, along the line-of-sight towards the observer. When a ray from the post-AGB star goes through the jet, the amount of absorption by the jet is calculated. The absorption is determined by the optical depth
\begin{equation}
    \Delta\tau_\nu(s) = c_\tau\, \rho_\text{sc}(s)\,\Delta s,\label{eq:tau_spatio}
\end{equation}
with $c_\tau$ the scaling parameter and $\Delta s$ the length of the line element at position $s(\theta,z)$. The scaled density $\rho_\text{sc}$ in this model is dimensionless and is a function of the polar angle $\theta$ and height $z$ in the jet, as follows:
\begin{equation}
    \rho_\text{sc}(\theta,z) = \left( \frac{\theta}{\theta_\text{out}} \right)^{p} \left( \frac{z}{\text{1\,AU}} \right)^{-2},
\end{equation}
with $p$ the exponent, which is a free parameter in the model, $\theta_\text{out}$ the outer jet angle, and $z$ the height of the jet above the centre of the jet cone. Hence, we calculate the relative density structure, which can then be scaled by the scaling parameter $c_\tau$, in order to fit the synthetic spectra to the observations. By doing so, the computations of optical depth are fast. The absolute density of the jet is estimated in Sect.~\ref{ssec:rtmodel}. 

In this model, we implement the same three jet configurations as in \cite{bollen19}: a stellar jet, an X-wind, and a disk wind. The velocity profile used for the stellar jet and X-wind models is defined as
\begin{equation}
    v(\theta) = v_\text{out} + \left( v_\text{in} - v_\text{out} \right) \cdot f_1(\theta), \label{eq:v_xwind}
\end{equation}
where $v_\text{out}$ and $v_\text{in}$ are the outer and inner velocities, and with
\begin{equation}
   f_1(\theta) = \frac{ e^{-p_\text{v}\cdot f_2(\theta)} - e^{-p_\text{v}}  }{1 - e^{-p_\text{v}}}.
\end{equation}
$p_\text{v}$ is a free parameter and $f_2$ is defined as
\begin{equation}
    f_2(\theta) = \left| \frac{ \theta - \theta_\text{cav} }{ \theta_\text{out} - \theta_\text{cav}} \right|,
\end{equation}
where $\theta_\text{out}$ is the outer jet angle and $\theta_\text{cav}$ the cavity angle of the jet.

The velocity profile of the disk wind is dependent on the Keplerian velocity at the location in the disk, from where the material is ejected. For the inner jet region between the jet cavity and the inner jet angle ($\theta_\text{cav} < \theta < \theta_\text{in}$), we have
\begin{equation}
    v(\theta) = v_\text{in,cav} + \left( v_\text{in, sc} - v_\text{in,cav} \right) \cdot \left( \frac{ \theta - \theta_\text{cav} }{ \theta_\text{in} - \theta_\text{cav}} \right)^{p_\text{v}},\label{eq:diskwind_in}
\end{equation}
with $v_\text{in,cav}$ the jet velocity at the cavity angle ($\theta_\text{cav}$) and $v_\text{in, sc}$ the jet velocity at the inner boundary angle ($v_\text{in}$). For the outer jet region ($\theta_\text{in} < \theta < \theta_\text{out}$), the velocity is defined by
\begin{equation}
    v(\theta) = \frac{v_M}{\sqrt{\tan\theta}},\label{eq:diskwind_out}
\end{equation}
with 
\begin{equation}
    v_M = v_\text{out, sc}\sqrt{\tan\theta_\text{out}}.
\end{equation}
We define the scaled inner velocity as $v_\text{in, sc}=v_M\cdot \left(\tan\theta_\text{in}\right)^{-1/2}$ and the scaled outer velocity as $ v_\text{out, sc} = c_\text{v}\cdot v_\text{out}$. $v_\text{out}$ is the outer jet velocity, which is equal to the Keplerian velocity at the launching point in the disk. The scaling factor $c_\text{v}$ can have values between 0 and 1. Hence, the disk wind velocity is smaller than or equal to the Keplerian velocity from its launching point in the disk.

In the three jet configurations, we included two important updates. The first update is the ability to model jets whose axis is tilted with respect to the direction perpendicular to the orbital plane. As can be seen in Fig.~\ref{fig:bd_dyn}, the absorption feature is not completely centred on the phase of superior conjunction, i.e. when the MS companion is between the post-AGB primary and the observer. This can be explained by a tilt in the jet, causing the absorption feature to be observed later in the orbital phase. A tilted jet in the binary system would lead to a precessing motion of the jet. This is not uncommon and has been previously observed in pre-planetary nebulae \citep{sahai05, yung11}. We implemented this jet tilt as an extra free parameter in our fitting routine.

The second update to our model presented in \cite{bollen19} is the introduction of a jet cavity for the X-wind and the disk wind configurations. In \cite{bollen19}, we showed that the density in the innermost region of the outflow is extremely low and thus barely contributing to the absorption. This is in agreement with the disk wind theory by \cite{blandford82} and the X-wind theory by \cite{shu94}. According to these theories, the disk material is launched at angles of $30\,\degr$ with respect to the jet axis, although farther from the launch point the angle can decrease substantially due to magnetic collimation. In our model, we allow some flexibility for the cavity angle parameter, by giving it a lower limit of $20\,\degr$. We will compare the new version of the spatio-kinematic model with the older version that does not include the cavity and tilt during our analysis in Sects.~\ref{ssec:spatioBD} \& \ref{ssec:spatioIRAS}.


\subsection{Radiative transfer model of the jet}\label{ssec:rtmodel}
The spatio-kinematic model is used as input for the radiative transfer model. Hence, the geometry, velocity, and scaled density structure are fixed with the values estimated from the spatio-kinematic model. By calculating the radiative transfer through the jet, the absolute jet densities can be determined, from which we can then calculate the jets' mass ejection rate. Here, we use the equivalent width (EW) of the Balmer lines to fit the model to the observations. The fitting parameters are the absolute jet densities and temperatures, instead of relative density differences throughout the jet. Hence, the optical depth calculations become more CPU-intensive.

\subsubsection{Radiative transfer}\label{ssec:rt}
In our radiative transfer code, we assume thermodynamic equilibrium and the jet medium to be isothermal. Hence, each line-of-sight through the jet to the disk of the star will have the same temperature. Here, we use the formal solution of the one-dimensional radiative transfer equation, where the source function $S_\nu$ is described by the Planck function $B_\nu$ \citep[see chapter 1]{rybicki}. For the incident intensity of the post-AGB star in the model, we use a synthetic stellar spectrum from \cite{coelho14}, which is chosen based on the parameters of the post-AGB star.

Using Boltzmann's equation, and by expressing the Einstein coefficients in terms of the oscillator strength $f_{12}$, the absorption coefficient $\alpha_\nu$ can be written as follows:

\begin{equation}
    \alpha_\nu(s) = \frac{\pi e^2}{m_e c}\phi_\nu n_l(s) f_{lu}\Big[1 - \mathrm{e}^{-\Delta E/kT}\Big].\label{eq:op2} 
\end{equation}
with $n_l$ and $n_u$ the densities in the lower and upper energy level, $f_{lu}$ the oscillator strength, and $\Delta E$ the energy difference between the upper and lower energy levels.
Hence, the computation of the intensity is dependent on the number density $n_l$, the temperature $T$, and the normalised line profile $\phi_\nu$. 
This normalised line profile $\phi_\nu$ is described as a Doppler profile for H$\beta$, H$\gamma$, and H$\delta$. For H$\alpha$, we follow the description in \cite{muzerolle01}, \cite{kurosawa06}, and \cite{kurosawa11} instead. As \cite{muzerolle01} showed, the Stark broadening effect can become significant in the optically thick H$\alpha$ line. Hence we describe the line profile of H$\alpha$ with the Voigt profile:
\begin{equation}
    \phi(\nu) = \frac{1}{\pi^{1/2}\Delta \nu_\mathrm{D}}\frac{a}{\pi}\int_{-\infty}^{\infty}\frac{\mathrm{e}^{-y^2}}{\left(\frac{\Delta\nu}{\Delta\nu_\mathrm{D}}-y \right)^2 + a^2} \dd y,
\end{equation}
with $\Delta \nu_\mathrm{D}$ the Doppler width of the line, $a=\Gamma/4\pi\Delta \nu_\mathrm{D}$, and $y = (\nu - \nu_0)/\Delta\nu_\mathrm{D}$ with $\nu_0$ the line centre. The Doppler line width $\Delta \nu_\mathrm{D}$ is a function of the thermal velocity $v_D$:
\begin{equation}
    \Delta \nu_\mathrm{D} = \nu_0 \frac{v_\mathrm{D}}{c}= \frac{\nu_0}{c}\sqrt{\frac{2kT}{m_p}}.
\end{equation}
We use the damping constant $\Gamma$ as described by \cite{vernazza73}, which is given by the sum of the natural broadening, Van der Waals broadening and the linear Stark broadening effects:
\begin{equation}
    \Gamma = C_\mathrm{Rad} + C_\mathrm{VdW} \Bigg( \frac{n_{HI}}{10^{22}\, \mathrm{m}^{-3}}\Bigg) \Bigg(\frac{T}{5000\,\mathrm{K}}\Bigg)^{0.3} + C_\mathrm{Stark} \Bigg( \frac{n_e}{10^{18}\,\mathrm{m}^{-3}}\Bigg)^{2/3}, \label{eq:damp}
\end{equation}
with $C_\mathrm{Rad}$, $C_\mathrm{VdW}$, and $C_\mathrm{Stark}$ the broadening constants of the natural broadening, Van der Waals, and Stark broadening effects, respectively, $n_{HI}$ the neutral hydrogen number density, and $n_e$ the electron number density. For the broadening constants, we use the values from  \cite{luttermoser92}: $C_\mathrm{rad}=8.2\times10^{-3}\angstrom$, $C_\mathrm{VdW}=5.5\times10^{-3}\angstrom$, and $C_\mathrm{Stark}=1.47\times10^{-2}\angstrom$.

\subsubsection{Numerical integration of the radiative transfer equation} \label{sssec:rtnum}

\begin{figure}[h!]
\centering
\includegraphics[width=.5\textwidth]{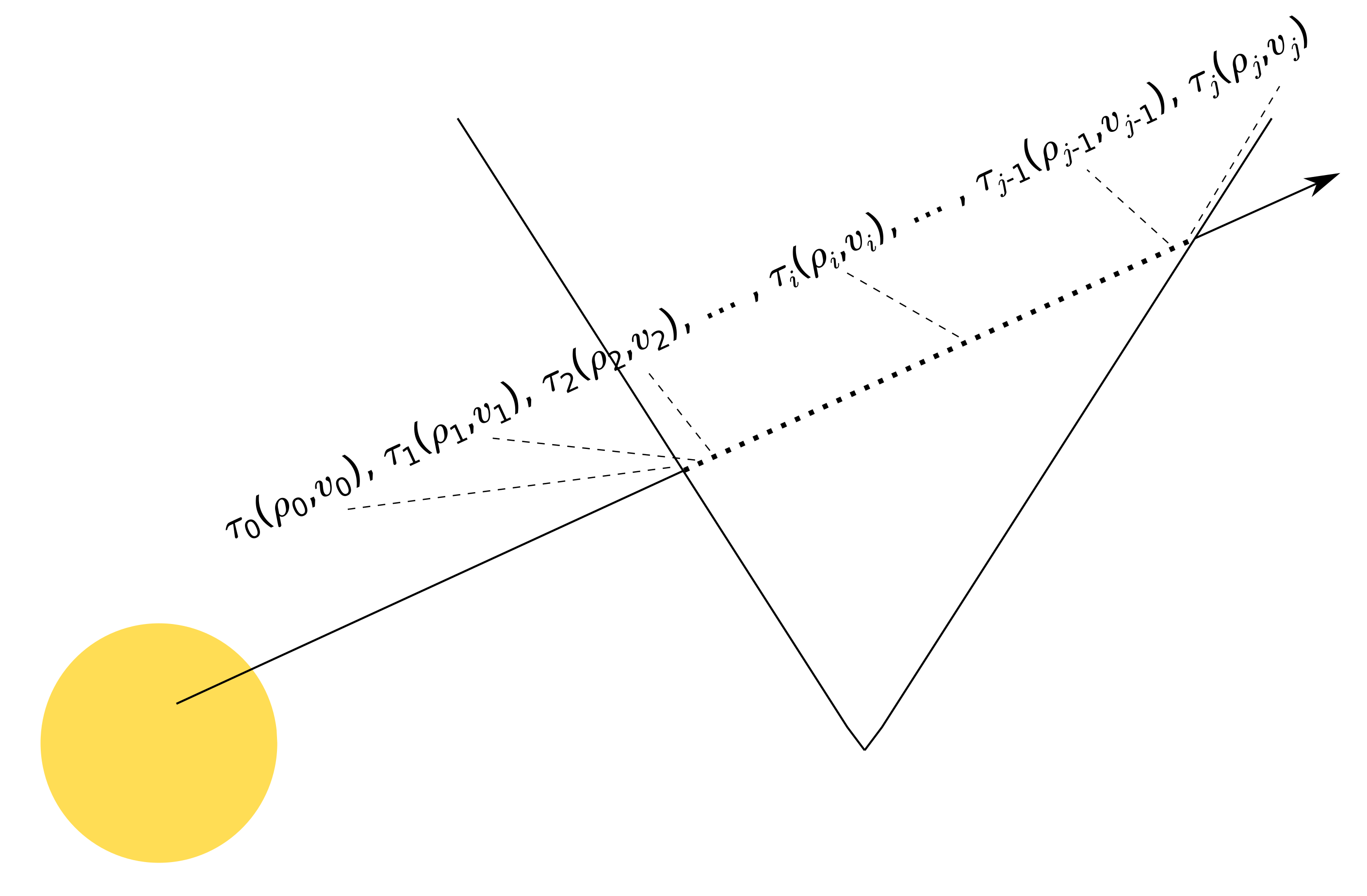}
\caption{A schematic representation of the radiative transfer calculations in the jet. The ray travelling from the star to the observer is split in $N_j$ grid points where it passes through the jet. Each grid point has a density $\rho$ and a velocity $v$. We iterate over each grid point in order to determine the resulting intensity along this line, for each wavelength.}\label{fig:schematicjet}
\end{figure}
In our model, we divide the light from the post-AGB star up in $N_r$ rays. To compute the radiative transfer through the jet, we solve the one-dimensional radiative transfer equation numerically. Hence, we iterate over each grid point along each ray, as shown in Fig. \ref{fig:schematicjet}. This ray is split up in $N_j$ grid points between the point of entry and exit in the jet. The intensity at a grid point $i$ along the ray is computed as follows:
\begin{equation}
     I_\nu(\tau_i) = I_\nu(\tau_{i-1}) \,\mathrm{e}^{-(\tau_{i-1}-\tau_{i})} + B_\nu(\tau_{i})\,\Big[1-\mathrm{e}^{-(\tau_{i-1} - \tau_{i})}\Big].
\end{equation}
 Hence, if we want to calculate the intensity through the whole line-of-sight through the jet, the observed intensity $I_\nu(\tau_{n})$ will be
\begin{align}
\begin{split}
    I_\nu(\tau_n) \,= \,&I_\nu(\tau_{0})\,\mathrm{e}^{-(\tau_{0}-\tau_{n})} + B_\nu(\tau_{n})\,\Big[1-\mathrm{e}^{-(\tau_{n-1} - \tau_{n})}\Big] \\
    &+ \sum\limits_{i=1}^{n-1} B_\nu(\tau_{i})\,\Big[1-\mathrm{e}^{-(\tau_{i-1} - \tau_{i})}\Big]\,\mathrm{e}^{-(\tau_{i}-\tau_{n})}. \label{eq:rtnum}
\end{split}
\end{align}
Since the ray has been divided into discrete intervals, the optical depth $\tau_i$ will be calculated as
\begin{equation}
    \tau_{i+1} - \tau_{i} = \rho_i \kappa_{\nu,i}\,\Delta s = \alpha_{\nu,i}\,\Delta s,\label{eq:optdepth_sum}
\end{equation}
with $\kappa_{\nu,i}$ the opacity.
This procedure is iterated for each ray from the post-AGB star and each frequency $\nu$. Hence, in general, the model consists of $N_r$ rays leaving the post-AGB surface, which are divided in $N_j$ grid points and for which the intensity is computed for a total of $N_\nu$ frequencies. As we are assuming that the jet is isothermal and the jet velocities significantly smaller than the speed of light ($v/c \ll 1$), we do not need to compute the Planck function $B_\nu$ for each grid point. However, this more general formulation will allow non-isothermal jet models to be computed in the future.
 
\subsubsection{Equivalent width as tracer of absorption}\label{ssec:obsew}
 
The photospheric light from the post-AGB star that travels through the jet will be scattered by the hydrogen atoms in the jet, causing the absorption features in the Balmer lines. To quantify this scattering in our model and the observations, we use the EW of the Balmer lines as fitting parameter. We do this for two main reasons. The first reason is that the EW of a line will be higher for stronger extinction. Hence, the EW quantifies the amount of scattering by the jet. The EW is highly dependent on the level populations of hydrogen at the location where the line-of-sight passes through the jet. In our model, these level populations are determined by the local density and temperature of the jet at those locations.

Second, the ratio of EW between the four Balmer lines, i.e. H$\alpha$, H$\beta$, H$\gamma$, and H$\delta$, is also dependent on the chosen jet temperature and density. This ratio can change dramatically when these two parameters are changed. This makes the EW an ideal quantity in our fitting to find the absolute jet densities and temperatures.

\section{Jet model for \bd.}\label{sec:jetmodelBD}
\begin{figure*}[h!]
\captionsetup{width=1.\textwidth}
\centering
  \begin{tabular}{@{}c@{}c}
  \includegraphics[width =.5\textwidth]{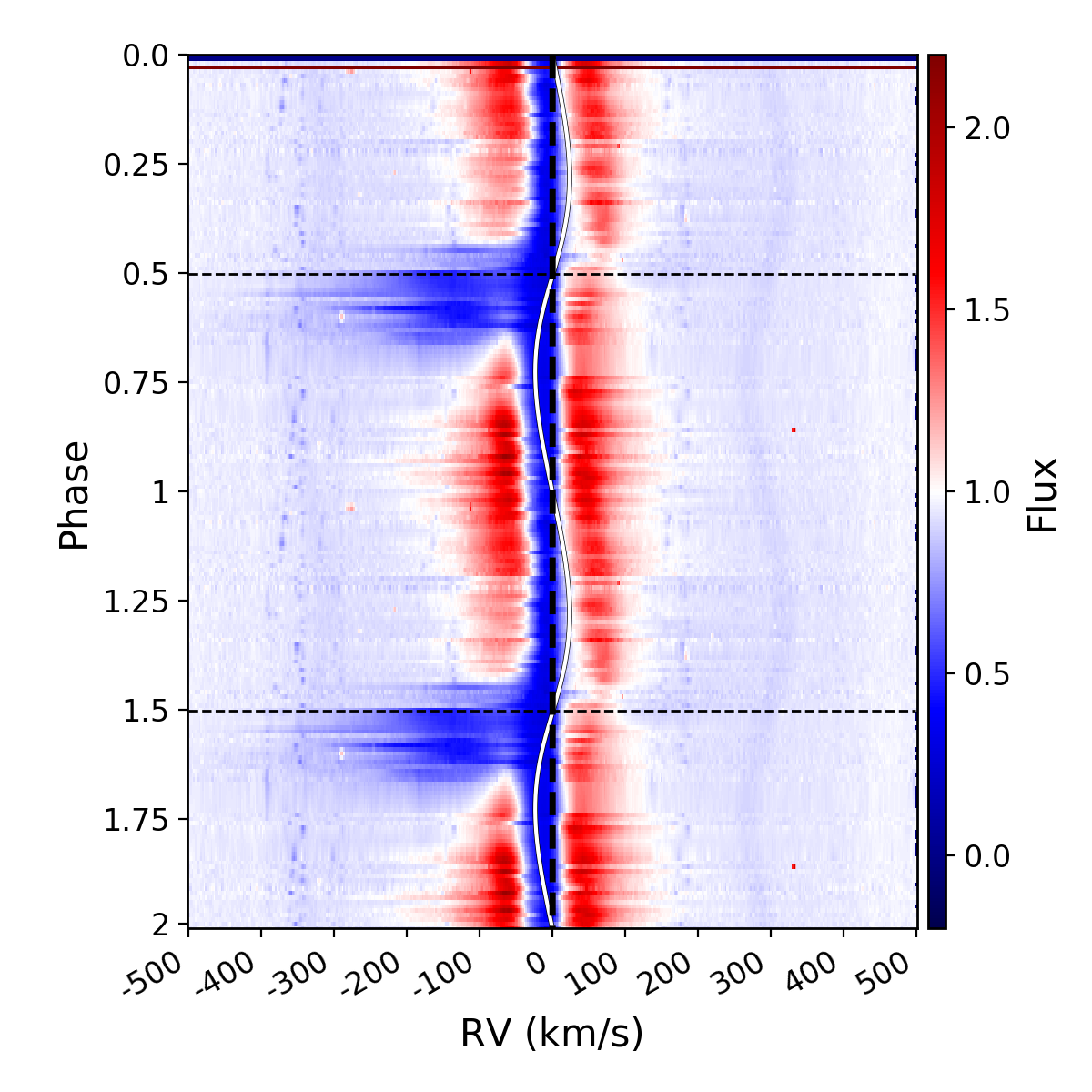}
&
	\includegraphics[width = .5\textwidth]{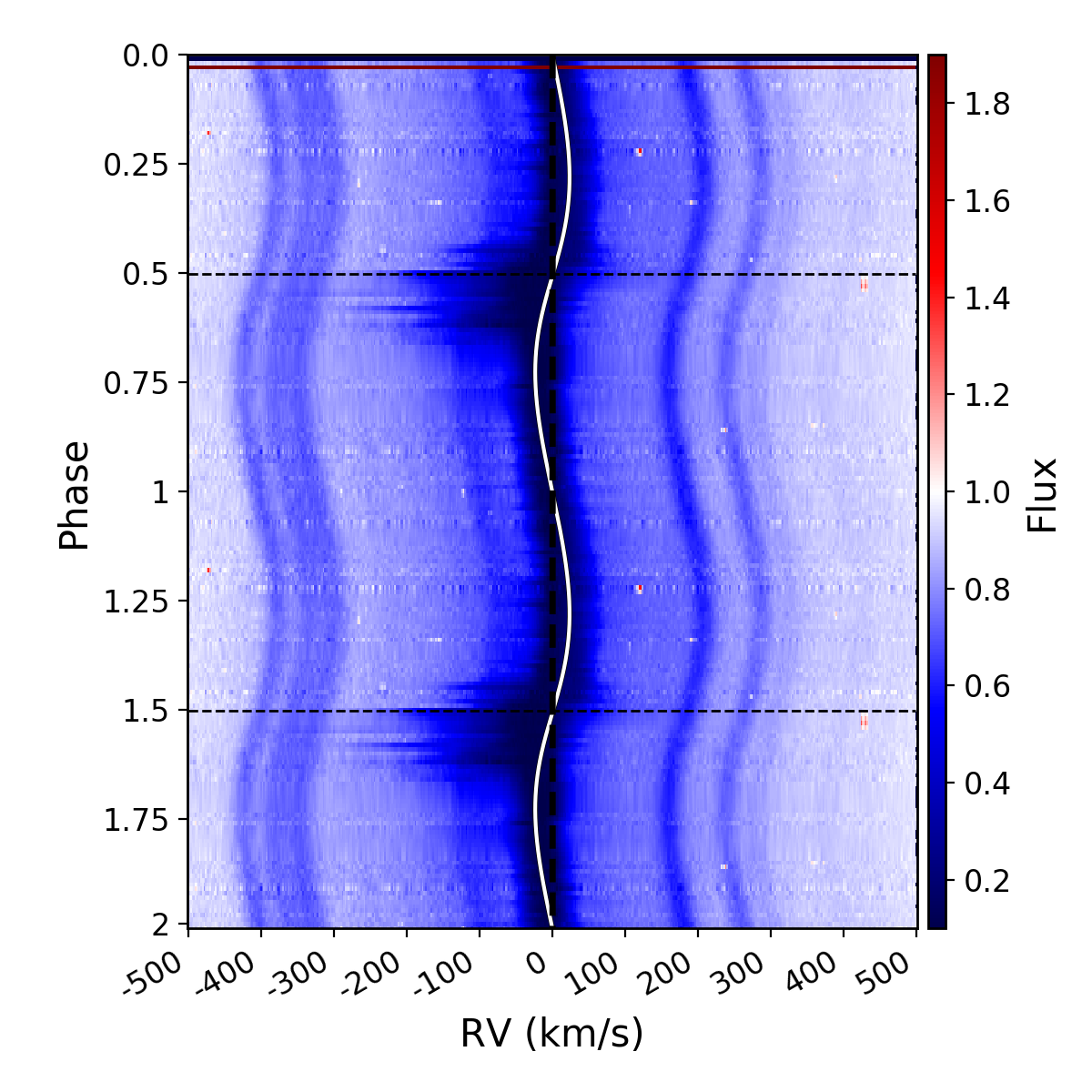} \\
  \includegraphics[width =.5\textwidth]{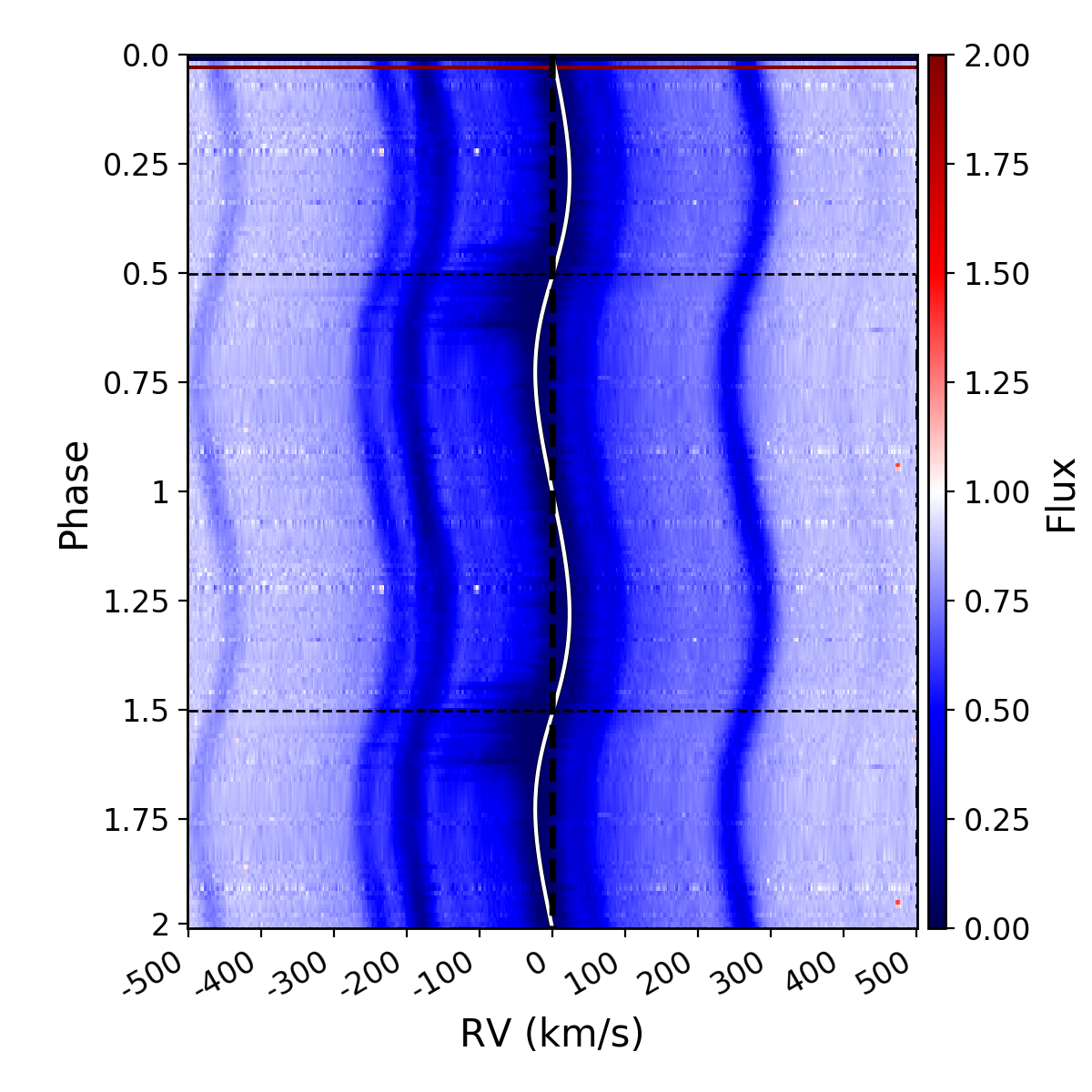}
&
	\includegraphics[width = .5\textwidth]{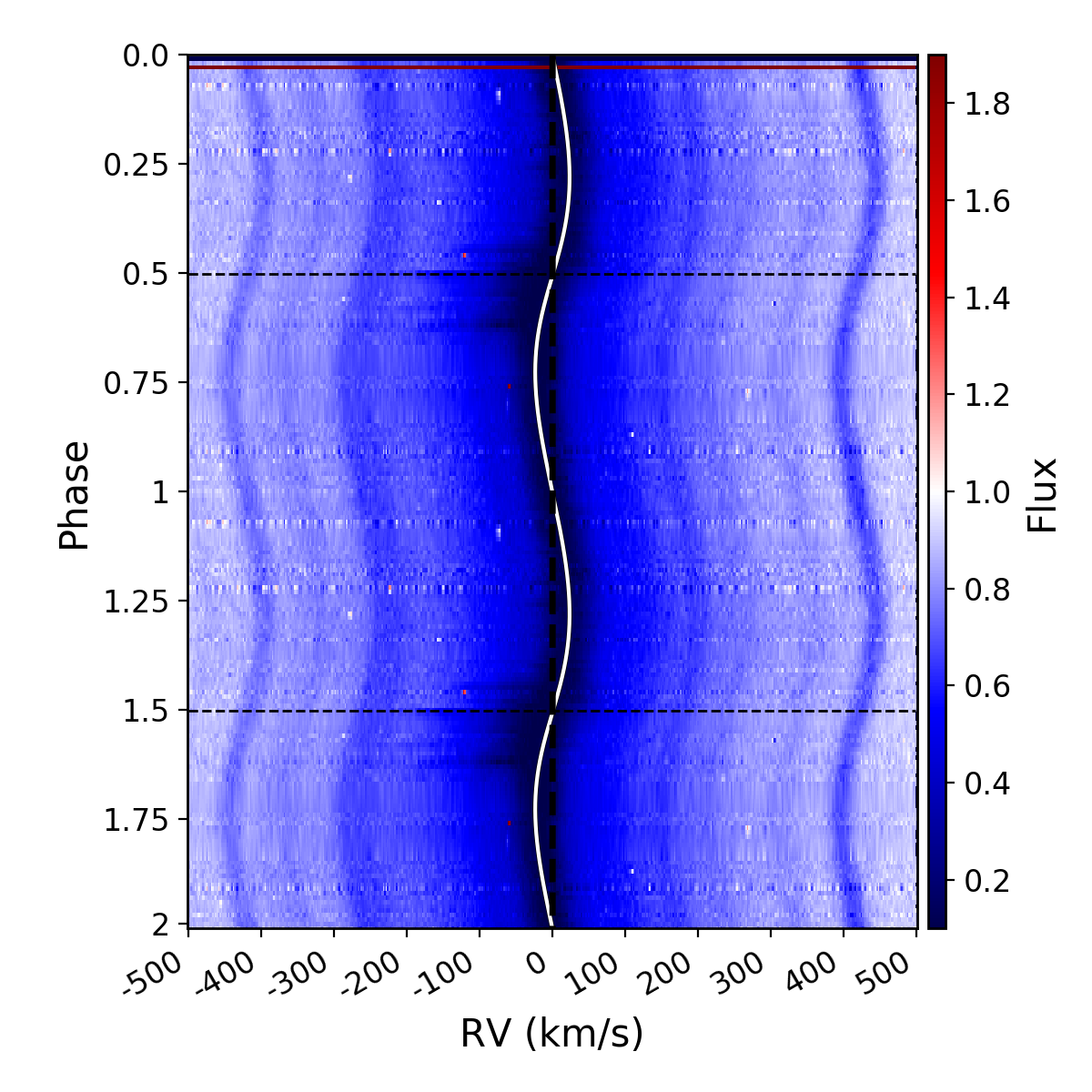} 
	
\end{tabular}
\caption{Dynamic spectra for the Balmer lines of \bd. Upper left: H$\alpha$, Upper right: H$\beta$, lower left: H$\gamma$, lower right: H$\delta$. The black dashed line indicated the phase of superior conjunction. The white line indicates the radial velocity of the post-AGB star. The colour gradient represents the strength of the line at a certain phase.}\label{fig:bd_dyn}

\end{figure*}

\bd\, is a jet-launching post-AGB binary system, for which we have obtained 36 spectra during one-and-a-half orbital cycles of the binary orbit of $140.82\,$days \citep{vanwinckel09, oomen18}. In this study, we adopt the orbital parameters listed in \citet[][see Appendix~\ref{ap:orbpar}]{oomen18}. The scattering by the jet is observable in the first four Balmer lines, e.g. H$\alpha$, H$\beta$, H$\gamma$, and H$\delta$, hence, we will focus on these line for our analysis. The Balmer lines are show in Appendix~\ref{ap:balmerlines}. The signal-to-noise ratio of the spectra lies between $\text{S/N}=22$ and $\text{S/N}=60$ in the H$\alpha$ line, and drops to values between $\text{S/N} = 12$ and $\text{S/N}=37$ in the H$\delta$ line. In Fig.~\ref{fig:bd_dyn}, we present the dynamic Balmer line spectra for \bd. In the dynamic spectra, we plot the continuum-normalised spectra as a function of orbital phase and interpolate between each of the spectra. In this way, the orbital phase-dependent variations in the line become apparent.

\subsection{Spatio-kinematic model of BD+46$^\circ$442}\label{ssec:spatioBD}

\begin{figure}[h!]
\centering
  \includegraphics[width =.5\textwidth]{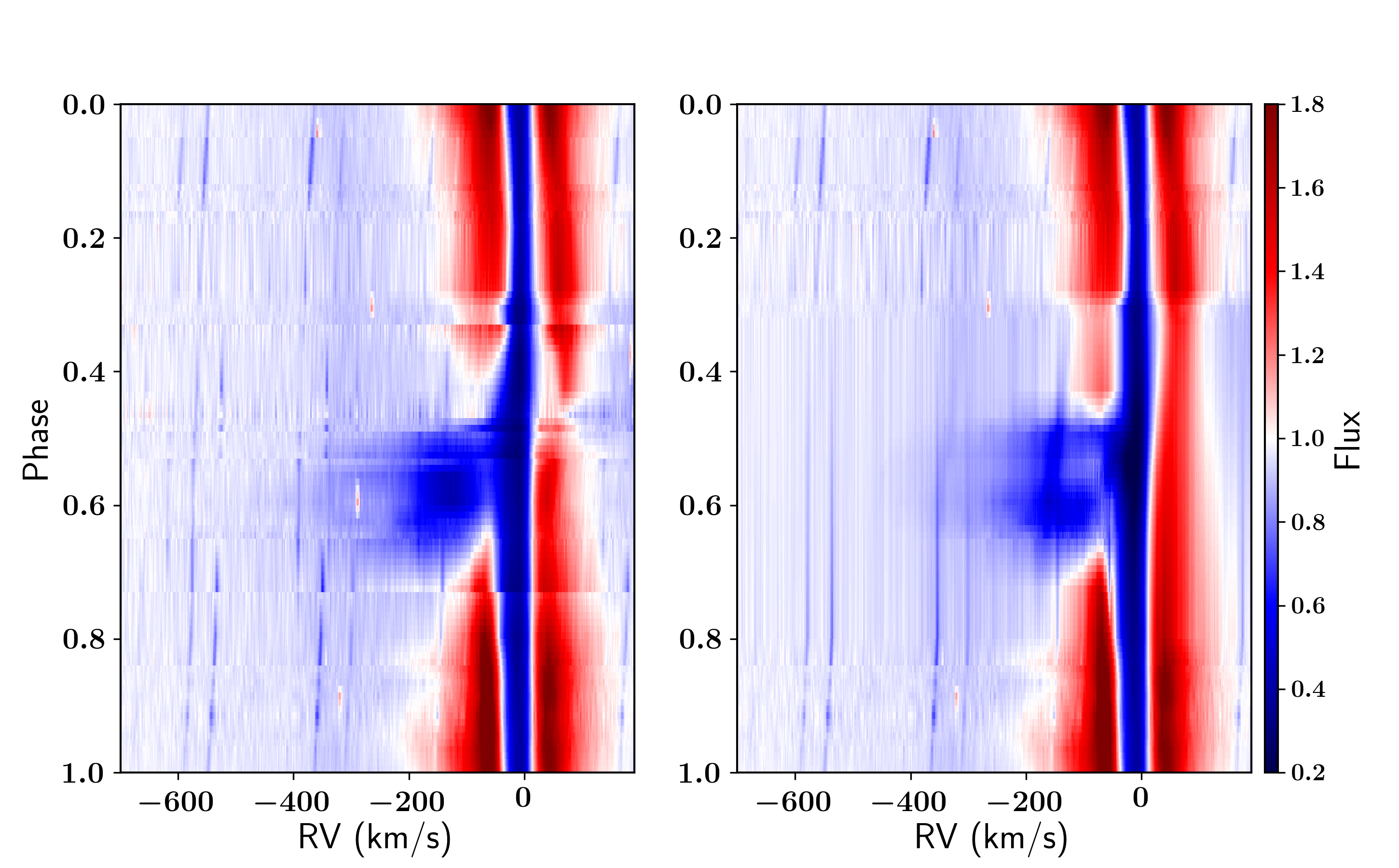}\\
	\includegraphics[width = .5\textwidth]{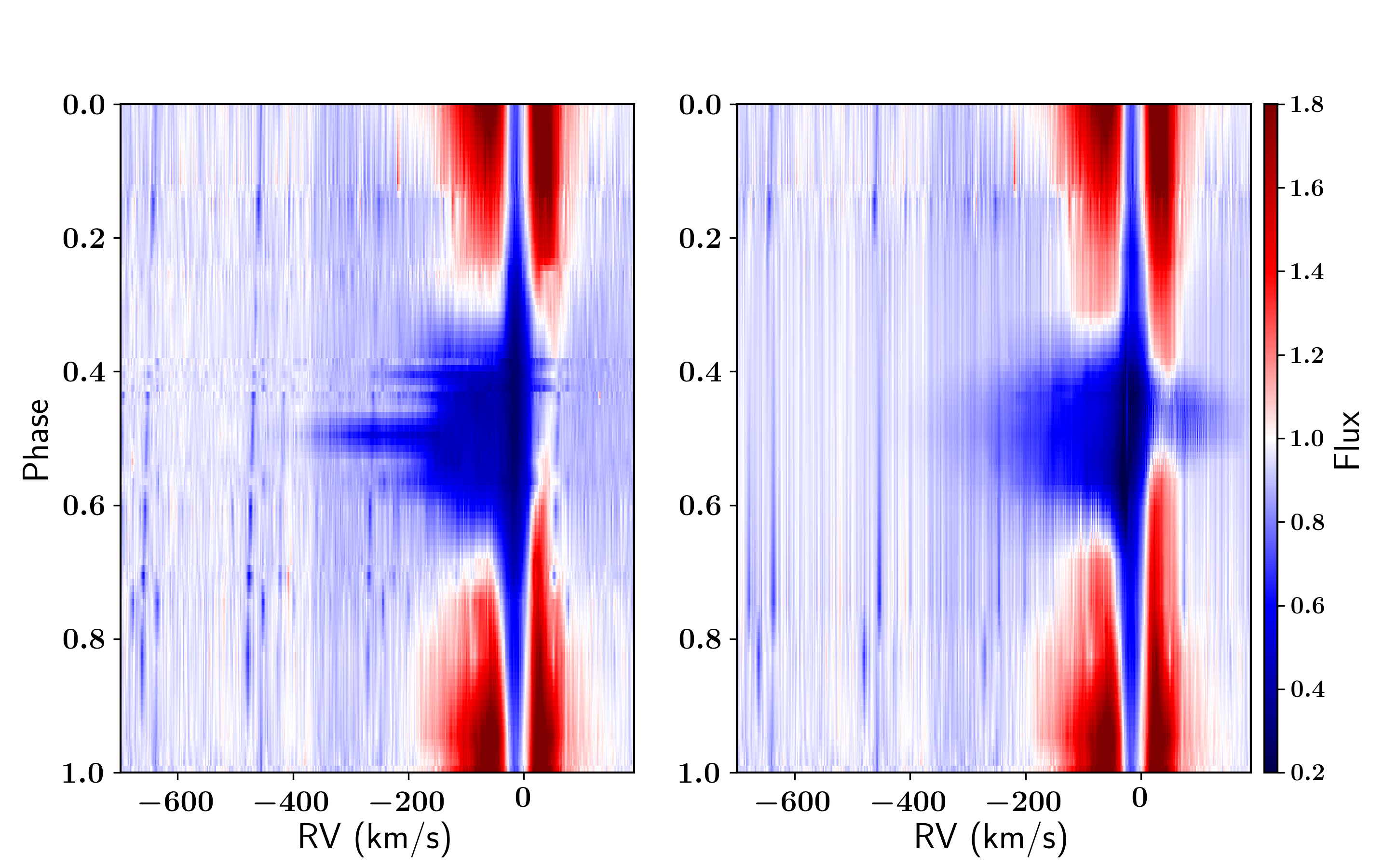}
\caption{Interpolated observed and modelled dynamic spectra of the H$\alpha$ line. The upper two spectra are the observations (left) and model spectra (right) of \bd. The lower two spectra are the observations and model spectra of \iras. The colours represent the normalised flux.}\label{fig:obsmod}
\end{figure}
\begin{figure*}
\centering
\includegraphics[width=1.\textwidth]{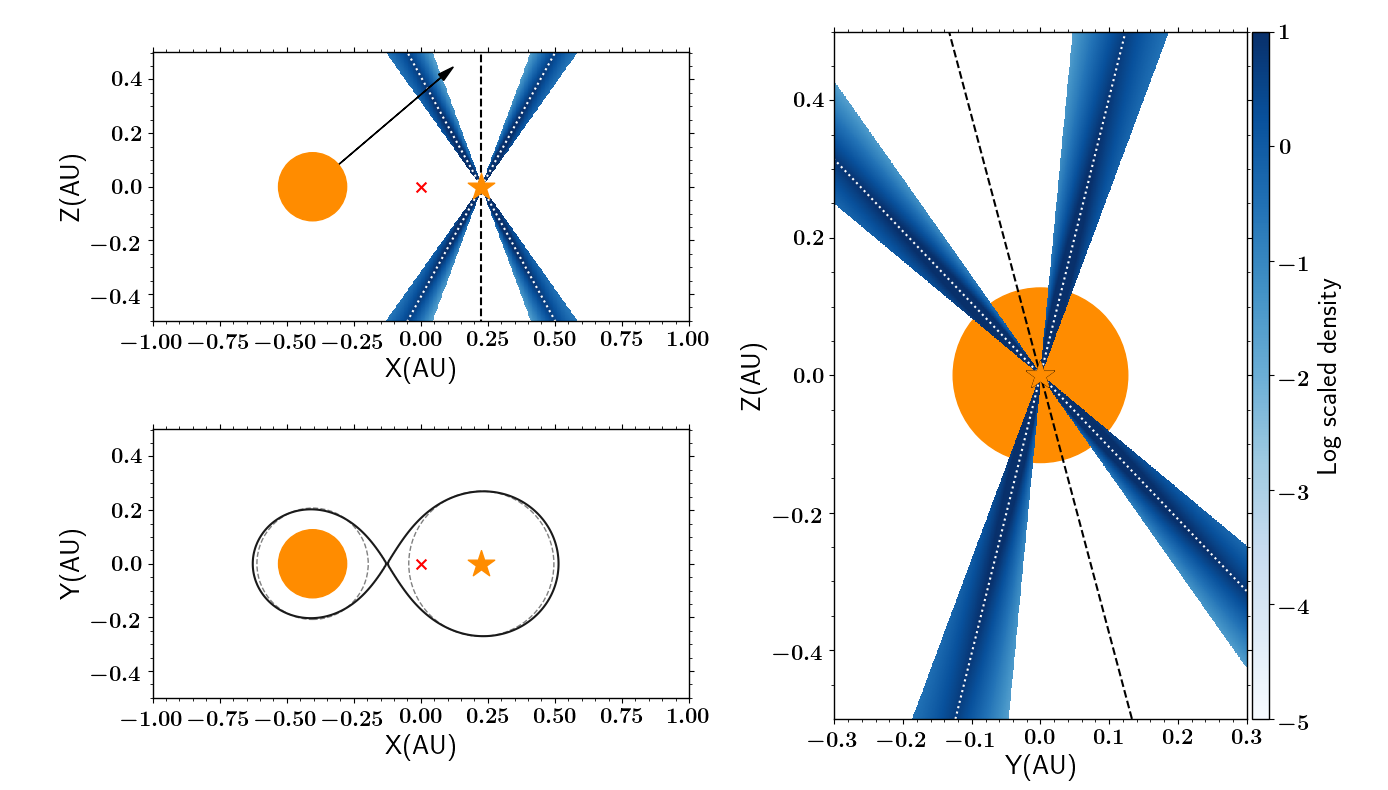}
\caption{The geometry of the binary system and the jet of \bd\, at superior conjunction, when the post-AGB star is directly behind the jet, as viewed by the observer. In all 3 plots, the full orange circle denotes the post-AGB star. The orange star indicates the location of the companion. The red cross is the location of the centre of mass of the binary system. The radius of the post-AGB star is to scale. The jet is represented in blue, where the colour indicates the relative density of the jet. The dashed black line is the jet axis and the dotted white lines are inner jet edges. The jet cavity is the inner region of the jet. The plot in the upper-left panel shows the system viewed along the orbital plane, from a direction perpendicular to the line-of-sight to the observer. The plot in the right panel shows the jet viewed from an angle perpendicular to the X-axis. The post-AGB star is located behind the companion and its jet in this image. The jet tilt is noticeable from this angle. The plot in the lower-left panel shows the binary system viewed from above, perpendicular to the orbital plane. The grey dashed lines represent the Roche radii of the two binary components and the full black line shows the Roche lobes. }\label{fig:bd_geometry}
\end{figure*}
\begin{table} 
\begin{center}
\caption{Best-fitting jet configuration and parameters for the spatio-kinematic model of \bd\, and \iras.}
\label{tab:bestfit}
\begin{tabular}{l c c}
\hline
\hline  \\
Parameter & \bd & \iras \\
\\
\hline \\
configuration                       & X-wind            & disk wind   \\
 $i$ ($\degr$)                      &   $49.6\pm0.6$    &  $71.8\pm0.5$     \\
 $\theta_\text{out}$ ($\degr$)      &   $35.3\pm0.5$    &  $67.3\pm1.3$   \\
 $\theta_\text{in}$ ($\degr$)       &   $28.8\pm0.5$    &  $64.4\pm2.5$     \\
 $\theta_\text{cav}$ ($\degr$)      &   $20.3\pm0.3$    &  $24\pm2.3$      \\
 $\phi_\text{tilt}$ ($\degr$)       &   $14.9\pm0.2$    & $5.7\pm1.3$      \\
 $v_\text{in}$ (km\,s$^{-1}$)     &    $490\pm30$     &  $640\pm40$   \\
 $v_\text{out}$ (km\,s$^{-1}$)      &   $41\pm5$        &  $66\pm6$     \\
 $c_\text{v}$                       &                   &   $0.044\pm0.065$    \\
 $p_\text{v}$                       &    $4.0\pm0.4$    &    $-0.8\pm1.4$   \\
 $p_{\rho\text{,out}}$              &   $-14.4\pm0.6$   &  $14.7\pm0.2$     \\
 $p_{\rho\text{,in}}$               &   $14.6\pm0.7$    &   $10\pm3$    \\
 $c_\tau$                           &   $2.08\pm0.06$   &  $1.51\pm0.14$  \\
 $R_\text{1}$ (AU)               &  $0.127\pm0.003$  &  $0.105\pm0.002$  \\ \\
\hline
\end{tabular}
\end{center}
\tablefoot{The tabulated parameters are: inclination angle of the binary system $i$, jet outer angle $\theta_\text{out}$, jet inner angle $\theta_\text{in}$, jet cavity angle  $\theta_\text{cav}$, jet tilt $\phi_\text{tilt}$ , inner jet velocity  $v_\text{in}$, jet velocity at the jet edges  $v_\text{out}$, velocity scaling parameter of the disk wind  $c_\text{v}$, exponent for the velocity profile  $p_\text{v}$, exponent for the density profile for the outer and inner region  $p_{\rho\text{,out}}$ and  $p_{\rho\text{,in}}$, optical depth scaling parameter  $c_\tau$, and the radius of the post-AGB star $R_\text{1}$.}
\end{table}
We compare the quality of the fit for the three jet configurations through their reduced chi-square and Bayesian Information Criteria (BIC) values\footnote{The BIC will penalise models that have a higher number of model parameters. Hence, the BIC is an ideal measure to compare the goodness of fit between our three jet configurations, since they do not have the same number of model parameters}. The best-fitting model is the X-wind with a reduced chi-square of $\chi^2_\nu=0.23$. A chi-square lower than unity indicates that the model is over-fitting the data. In our case, this is caused by overestimating the uncertainty on the data, which is determined from the signal-to-noise of the spectra ($\sigma =  (S/N)^{-1}$) and the uncertainty in the emission feature of the synthetic spectra that is provided as input for the modelling. We impose a chi square of unity for the best-fitting model and scale the $\chi^2$ of the other models appropriately, in order to compare their relative difference. The values of the scaled chi-square are: $\chi^2_\text{stellar}=1.12$, $\chi^2_\text{X-wind}=1$, and $\chi^2_\text{disk wind}=1.17$. Hence, the X-wind configuration gives a slightly better fitting result compared to the other two configurations. This is also confirmed by the BIC-values of the three models. The X-wind has the lowest BIC and therefore fits the data best: $\text{BIC}_\text{stellar} - \text{BIC}_\text{X-wind} = 1007$ and  $\text{BIC}_\text{disk wind} - \text{BIC}_\text{X-wind} = 1452$. For this reason, we will use the best-fitting parameters from the spatio-kinematic modelling of the X-wind for further calculations. We do note, however, that the relative difference in $\chi^2$ between the three model configurations is not significant, and thus, we conclude that the three model configurations fit the data equally well. 

The best-fitting parameters of the model are tabulated in Table~\ref{tab:bestfit} and its model spectra are shown in the upper right panel of Fig.~\ref{fig:obsmod}. The binary inclination for this model is about $50\,\degr$. The jet has a half-opening angle of $35\,\degr$. The inner boundary angle $\theta_\text{in}=29\,\degr$ is the polar angle in the jet along which the bulk of the mass will be ejected. The geometry of the binary system and the jet are represented in Fig.~\ref{fig:bd_geometry}. The material that is ejected in the inner regions of the jet reaches velocities up to $490\,$km s$^{-1}$. These velocities are of the order of the escape velocity from the surface of a MS star, confirming the nature of the companion. The velocities at the outer edges are lower at $41\,$km s$^{-1}$. The radius of the post-AGB star in the best-fitting model is $27.2$~\rsun ($0.127\,$AU).

Additionally, we implemented a jet tilt and jet cavity in this model (see Sect.~\ref{ssec:spatio}). The resulting model has a cavity angle of $20\,\degr$. The jet tilt for \bd\, is relatively large with $\phi_\text{tilt}= 15\,\degr$. The effect of this tilt is noticeable in the resulting model spectra. The jet absorption feature is not centred at orbital phase $0.5$, but at a later phase between $0.55-0.6$. To evaluate the performance of the new modified spatio-kinematic model that includes a jet cavity and tilt, we do an additional model-fitting with the old version that does not include these features and compare the model-fitting results between these two versions. The $\chi^2$ of the new model ($\chi^2_\text{new} = 1$) is lower than the old version ($\chi^2_\text{old}= 1.35$). If we account for the extra 2 parameters in the new model by comparing the BIC instead, we get a difference in BIC between the two results of $\Delta\,\text{BIC}=2980$, with the lower BIC for the new model, implying a better fit for this model. This demonstrates that the implemented jet cavity and jet tilt improve the spatio-kinematic model for this object.

\subsection{Radiative transfer model of \bd}\label{ssec:rtBD}
We apply the radiative transfer model for \bd\, to compute the amount of absorption caused by the jet that blocks the light from the post-AGB star. The setup for the radiative transfer model is similar to the one described in Sect. \ref{ssec:spatio}. For each ray of light, the background intensity $I_{\nu,0}$ is the background spectrum given in Sect. \ref{ssec:spatioBD}. for each orbital phase, we calculate the amount of absorption by the jet for each ray. Additionally, the output from the MCMC-fitting routine of Sect. \ref{ssec:spatioBD}, i.e. the spatio-kinematic model of \bd, is used as input in our the radiative transfer model. Hence, there are only two fitting parameters: jet number density $n_j$ and jet temperature $T_j$. We assume the jet temperature to be uniform for the segment of the jet through which the rays travel. The jet number density $n_j$ is defined as the number density at the inner edge of the jet $\theta_\mathrm{in}$ at a height of $1\,$AU. In the case of \bd, the best-fitting jet configuration is an X-wind. Hence the density in the jet at each grid point can be determined from the density profile of the jet that was used for the X-wind in the spatio-kinematic model. This density profile is defined as
\begin{equation}
    n(\theta,z) = n_j\, \left(\frac{\theta}{\theta_\mathrm{in}}\right)^{p} \, z^{-2},  \label{eq:densitystructure}
\end{equation}
with $p$ either the exponent for the inner-jet region $p_\text{in}$ or outer-jet region $p_\text{out}$, which was determined in Sect.~\ref{ssec:spatioBD}.

We use a grid of jet temperatures between 4400\,K and 6000\,K in steps of 100\,K and the logarithm of the jet densities $\log_{10}\left(\frac{n_j}{\mathrm{m}^{-3}}\right)$ between 14 and 18 in logarithmic steps of 0.1. This makes a total of 697 grid calculations.  

\begin{figure}
\centering
\includegraphics[width=.5\textwidth]{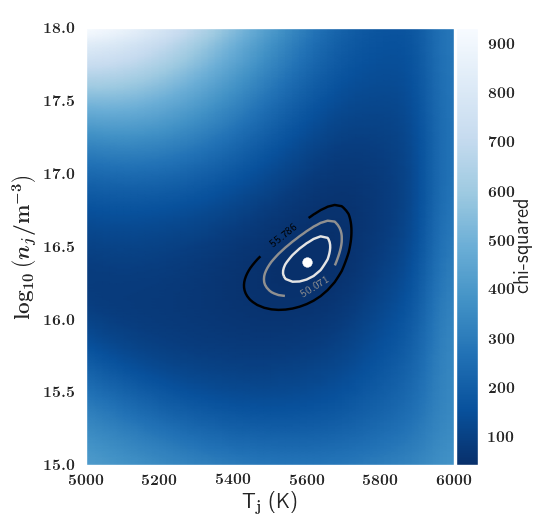}
\caption{2D reduced chi-squared distribution for the grid of jet densities $n_j$ and temperatures T$_j$ for the fitting of \bd. The white dot specifies the location of the minimum reduced chi-square value $\chi^2_{\nu,\text{min}} = 43.9$. The contours represent the 1-, 2-, and 3-$\sigma$ intervals.
}\label{fig:chisquared}
\end{figure}

As described in Sect.~\ref{ssec:obsew}, the EW of the Balmer lines represents the amount of absorption by the jet. In order to find the best-fitting model for our grid of temperatures and densities, we will fit the EW of the Balmer lines in the model to those of the observed Balmer lines for each spectra. 

We fit the model to the data with a $\chi^2$-goodness-of-fit test. Hence, the reduced $\chi^2_\nu$ value for a model will be
\small\begin{align}
\begin{split}
    \chi^2_\nu &= \frac{1}{\nu}\Bigg( \sum_i^{N_o}\left[\frac{\big(EW^{\,o,\,\text{H}\alpha}_i - EW^{\,m,\text{H}\alpha}_i\big)^2}{\big(\sigma^{\text{H}\alpha}_{i}\big)^2}\right] 
    + \sum_i^{N_o}\left[\frac{\big(EW^{\,o,\,\text{H}\beta}_i - EW^{\,m,\text{H}\beta}_i\big)^2}{\big(\sigma^{\text{H}\beta}_{i}\big)^2}\right] \\
    &+ \sum_i^{N_o}\left[\frac{\big(EW^{\,o,\,\text{H}\gamma}_i - EW^{\,m,\text{H}\gamma}_i\big)^2}{\big(\sigma^{\text{H}\gamma}_{i}\big)^2}\right] 
    + \sum_i^{N_o}\left[\frac{\big(EW^{\,o,\,\text{H}\delta}_i - EW^{\,m,\text{H}\delta}_i\big)^2}{\big(\sigma^{\text{H}\delta}_{i}\big)^2}\right]\Bigg),\label{eq:chi2}
    \end{split}
\end{align}
with $\nu$ the degrees of freedom, $N_o$ the number of spectra (36 for \bd, and 22 for \iras), $EW^o$ and $EW^m$ the equivalent width of the observed and modelled line, and $\sigma$ the standard deviation, which is determined by the signal-to-noise ratio of the spectra.

The resulting 2D $\chi^2_\nu$-\,distribution for jet densities and temperatures is shown in Fig.~\ref{fig:chisquared}. The best-fitting model has a jet density of $n_\text{j}=2.5\,\times\,10^{16}\,\text{m}^{-3}$ and jet temperature of $T_\text{j} = 5600\,$K. In order to determine the uncertainties on the fitting parameters, we convert the 2D chi-squared distribution into a probability distribution
\begin{equation}
    P_\text{2D}(n_j,T_j) \propto \exp\left( -\chi^2/2  \right). \label{eq:prob2d}
\end{equation}
The marginalised probability distribution can be found for each parameter by
\begin{align}
    P_\text{1D}(n_j) = \sum_{T_j} P_\text{2D}(n_j,T_j) \label{eq:prob1d_rho}\\
    P_\text{1D}(T_j) = \sum_{n_j} P_\text{2D}(n_j,T_j). \label{eq:prob1d_T}
\end{align}
From these distributions, we can determine a mean and standard deviation. This gives us a jet density and temperature of $n_j = 2.5^{+0.9}_{-0.7}\,\times\,10^{16}\,\text{m}^{-3}$ and $T_j = 5600\pm80\,$K.\footnote{The given uncertainties are the 2-$\sigma$ interval.}
\begin{figure}
\centering
\includegraphics[width=.5\textwidth]{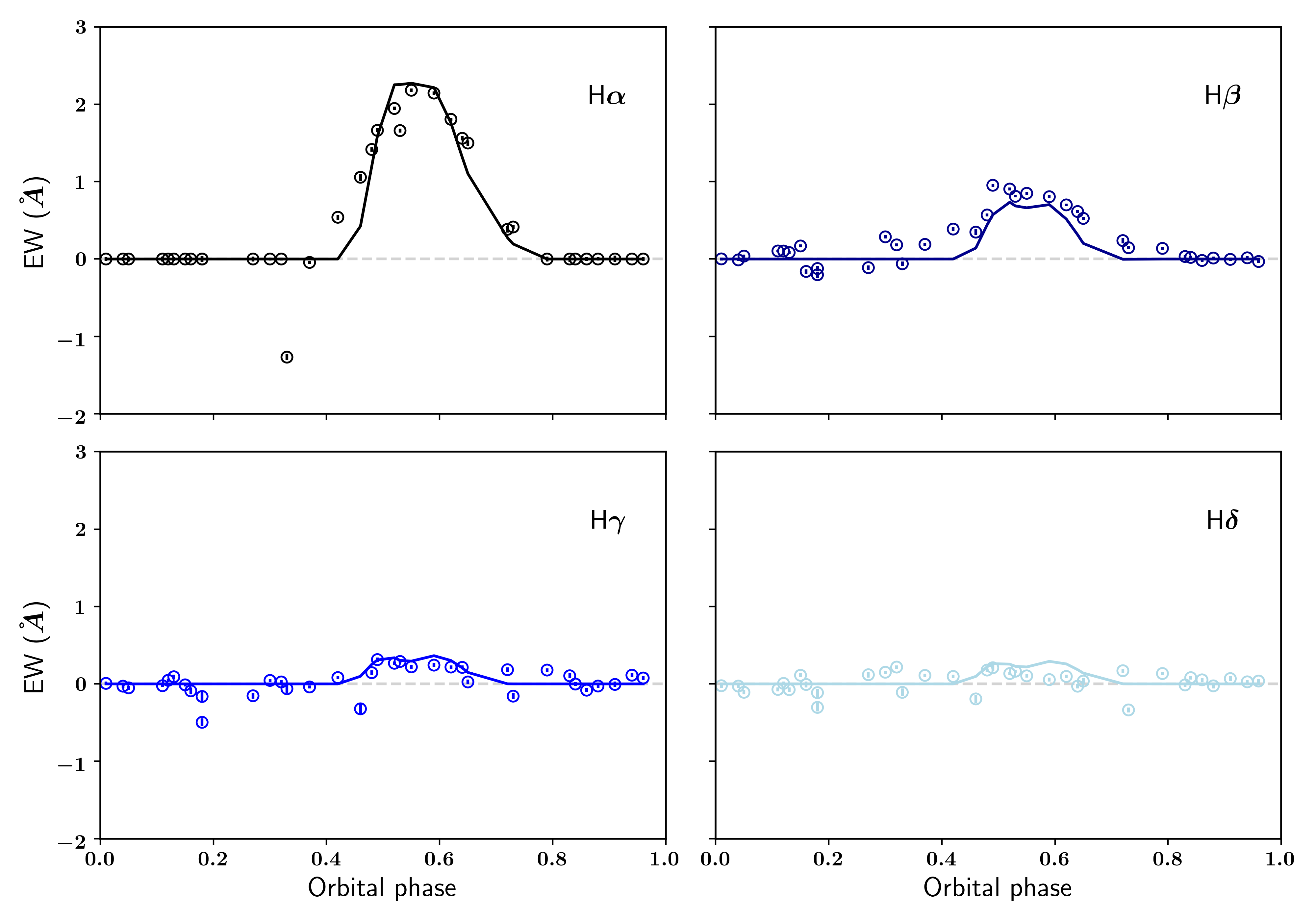}
\caption{The EW of the absorption by the jet for \bd\, as a function of orbital phase. The four panels show the EW in H$\alpha$ (top-left), H$\beta$ (top-right), H$\gamma$ (bottom-left), and H$\delta$ (bottom-right). The circles are the measured EWs of the absorption feature by the jet in the observations with their respective errors. The full line is the EW of the absorption feature for the best-fitting model.}\label{fig:bd_ewfit}
\end{figure}

\section{Jet model for \iras}\label{sec:jetmodelIRAS}
The second post-AGB binary system that we model is \iras. For this object, we have obtained 22 spectra during a full cycle with an orbital period of $126.97\,$days \citep{vanwinckel09, oomen18}. As for \bd, we adopt the orbital parameters of this system found by \cite{oomen18}. The individual and dynamic spectra are shown in Appendix~\ref{ap:balmerlines} and Fig.~\ref{fig:iras_dyn}, respectively. The signal-to-noise ratio for these spectra lie between $\text{S/N}=26$ and $\text{S/N}=49$ in H$\alpha$. In H$\delta$, the range of signal-to-noise is between $\text{S/N}=5$ and $\text{S/N}=20$.

\begin{figure*}[h!]
\captionsetup{width=1.\textwidth}
\centering
  \begin{tabular}{@{}c@{}c}
  \includegraphics[width =.5\textwidth]{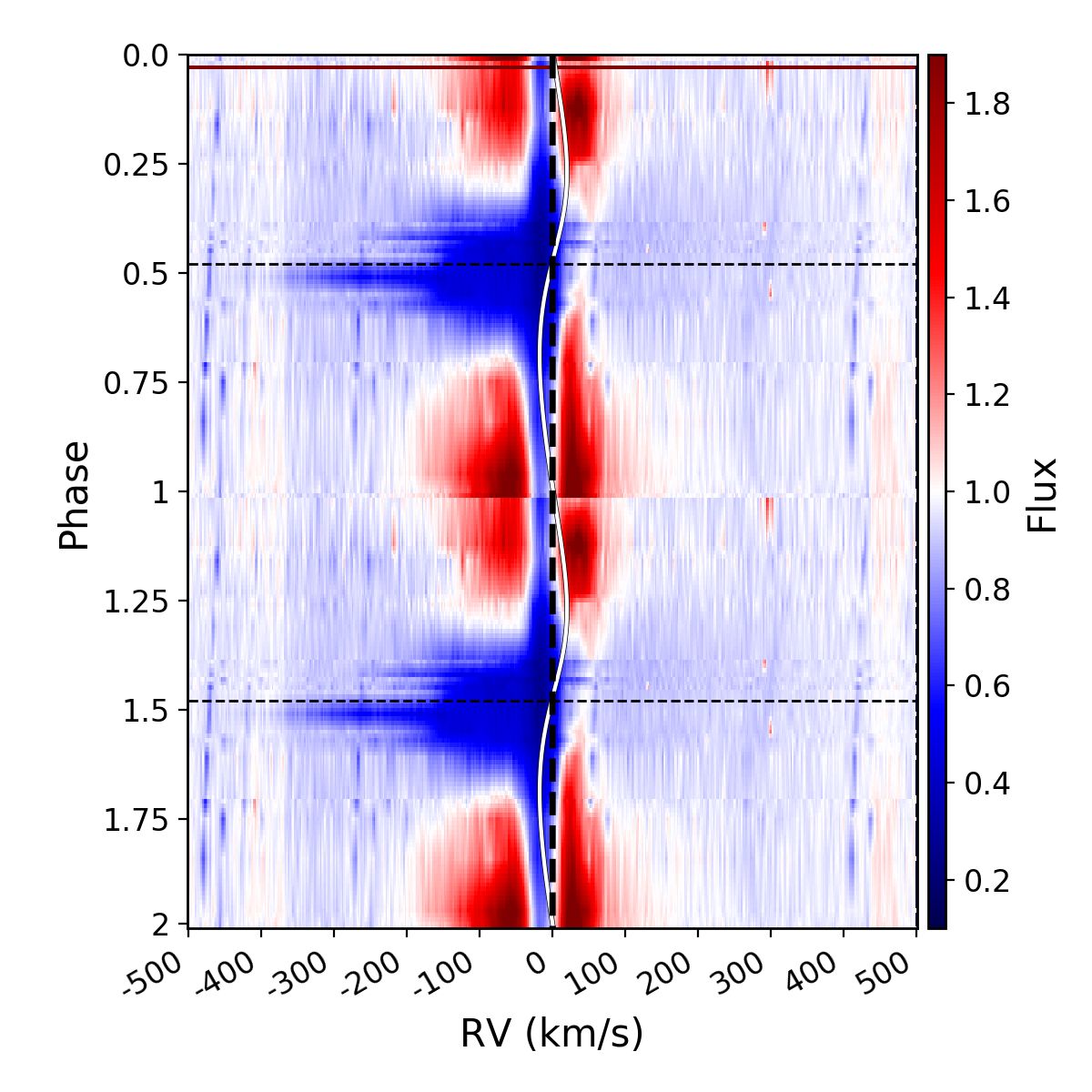}
&
	\includegraphics[width = .5\textwidth]{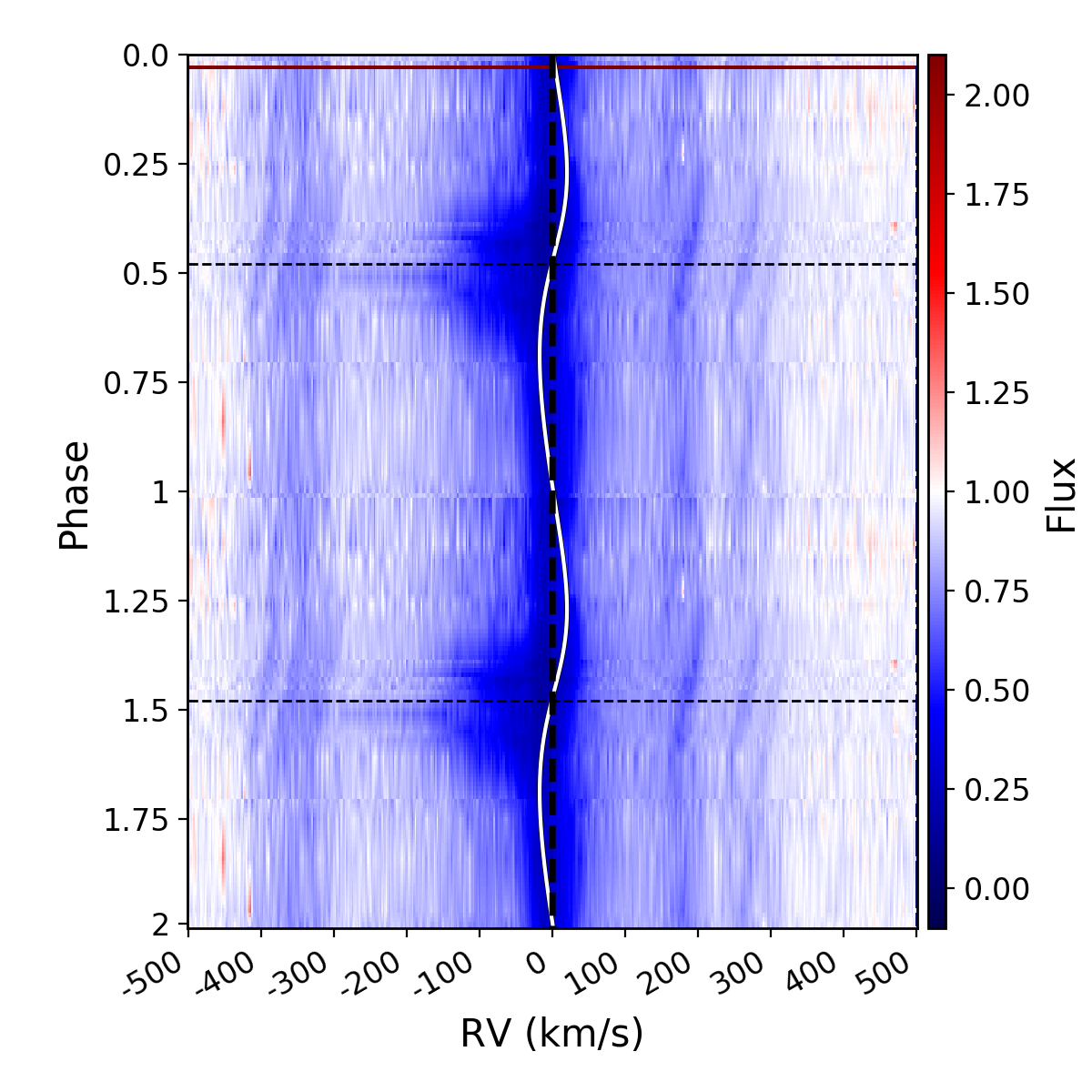} \\
  \includegraphics[width =.5\textwidth]{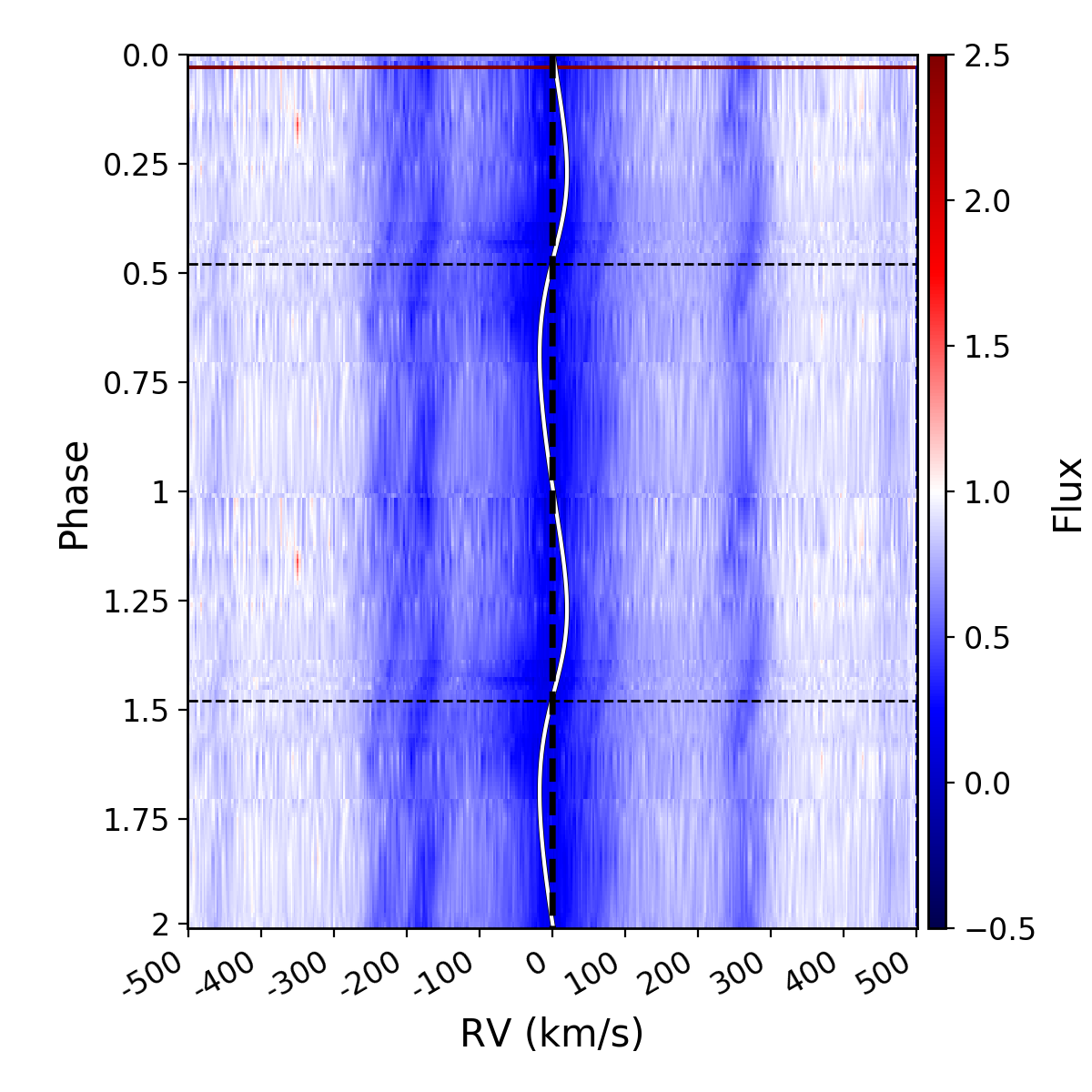}
&
	\includegraphics[width = .5\textwidth]{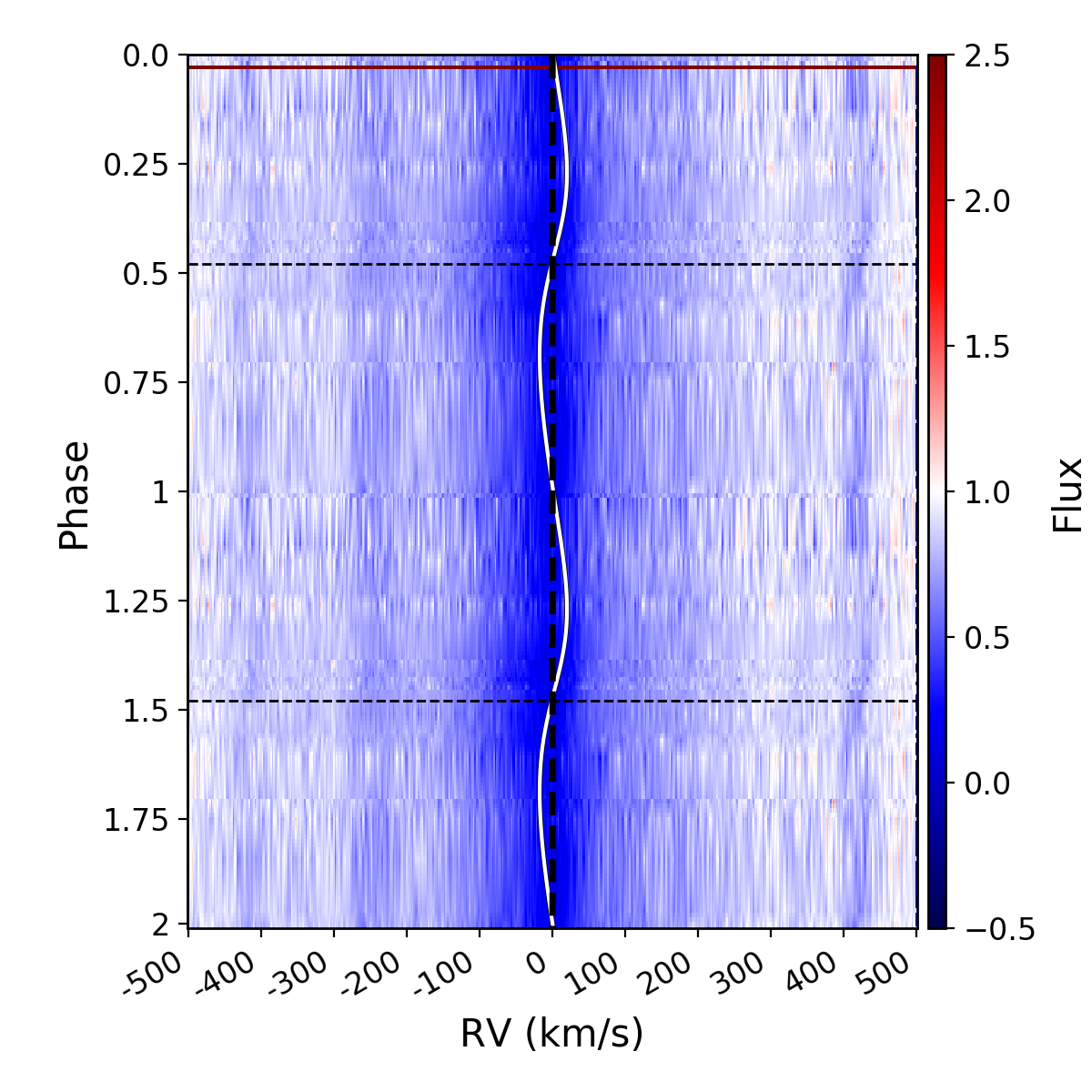} 
	
\end{tabular}
\caption{Dynamic spectra for the Balmer lines of \iras. Upper left: H$\alpha$, Upper right: H$\beta$, lower left: H$\gamma$, lower right: H$\delta$. The black dashed line indicated the phase of superior conjunction. The white line indicates the radial velocity of the post-AGB star. The colour gradient represents the strength of the line at a certain phase.}\label{fig:iras_dyn}
\end{figure*}

\subsection{Spatio-kinematic model of \iras}\label{ssec:spatioIRAS}

\begin{figure*}
\centering
\includegraphics[width=1\textwidth]{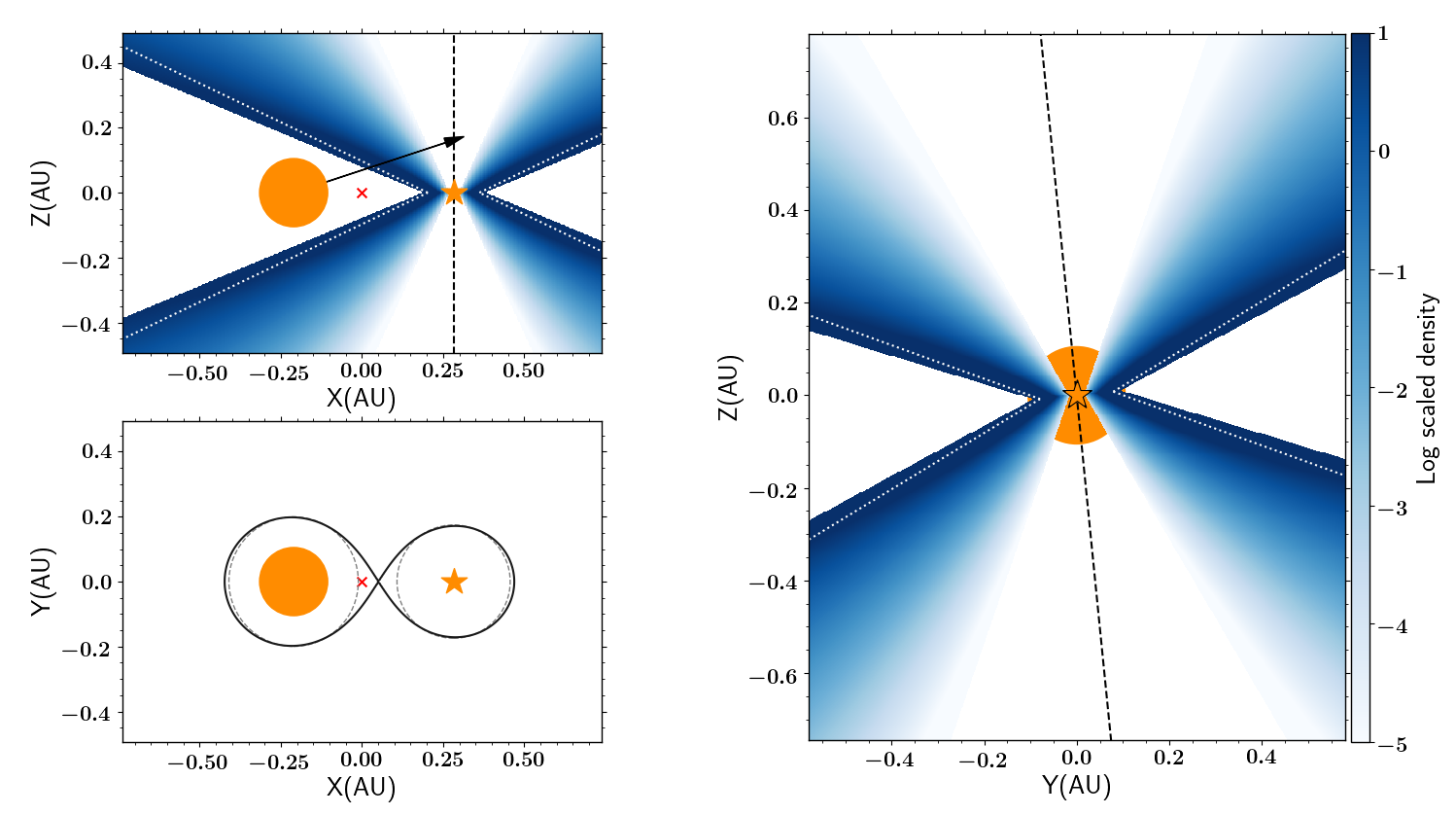}
\caption{Similar to Fig.~\ref{fig:bd_geometry}, but for \iras. }\label{fig:iras_geometry}
\end{figure*}
The spatio-kinematic structure of \iras\, has been modelled by \cite{bollen19}. Here, we update it with the addition of the jet tilt and the jet cavity. The best-fitting jet configuration is a disk wind, but all three models produce similar fits, as was the case in the fitting of \citet{bollen19} ($\text{BIC}_\text{stellar} - \text{BIC}_\text{disk wind} = 87$ and $\text{BIC}_\text{X-wind} - \text{BIC}_\text{disk wind} = 116$). The best-fitting model parameters are tabulated in Table~\ref{tab:bestfit}. This model has an inclination angle of $i = 72\,\degr$ for the binary system and a jet angle of $\theta_\text{out} = 67\,\degr$. These angles are about $7\,\degr$ lower than those found in the model-fitting of \cite{bollen19}. The jet reaches velocities up to $640\,$km s$^{-1}$. At its edges, the jet velocity is $v_\text{out}\cdot c_\text{v}=3\,$km s$^{-1}$. The post-AGB star in our model has a radius of $22.5$~\rsun ($0.105$~AU), which is about 30\% smaller than found by \cite{bollen19}. The geometry of the binary system and the jet are shown in Fig.~\ref{fig:iras_geometry}. We compare the quality of the fit for the best-fitting model of \cite{bollen19} with the best-fitting model in this work. The BIC for the model-fitting in our work is significantly lower than the BIC found by \cite{bollen19} ($\Delta \text{BIC}_\text{old} - \text{BIC}_\text{new}=4190$). This shows that the jet tilt and jet cavity significantly improve the model fitting. The jet tilt for this object is relatively small ($\phi_\text{tilt}=5.7\,\degr$). This is expected, since there is no noticeable lag in the absorption feature in the spectra. The jet for this object has a significant jet cavity of $\theta_\text{cav}=24\,\degr$.

\subsection{Radiative transfer model of IRAS19135+3937}\label{ssec:rtIRAS}
We apply the radiative transfer model for \iras. The best-fitting spatio-kinematic model found in Sect.~\ref{ssec:spatioIRAS} for \iras\, is a disk wind. We use this spatio-kinematic model and its model parameters as input to calculate the radiative transfer in the jet for a grid of jet densities and temperatures. The density profile for the disk wind is similar to the X-wind (see Eq.~(\ref{eq:densitystructure})). The grid of temperatures and densities is the same as for \bd.


\begin{figure}[h!]
\centering
\includegraphics[width=.5\textwidth]{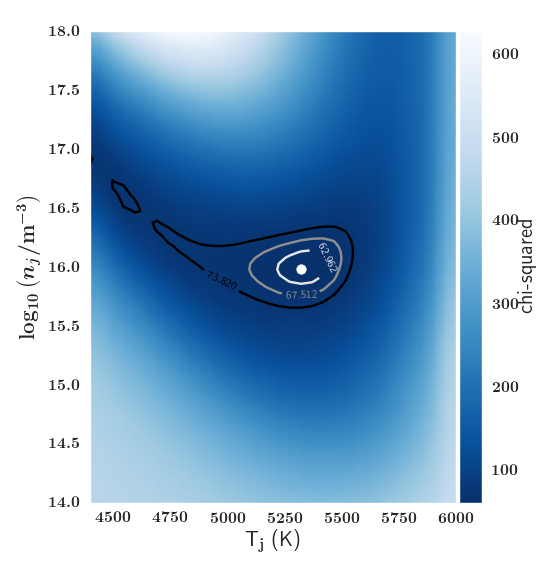}
\caption{2D reduced chi-squared distribution for the grid of jet densities $n_j$ and temperatures T$_j$ for the fitting of \iras. The white dot specifies the location of the minimum reduced chi-square value $\chi^2_{\nu,\text{min}} = 60.8$. The contours represent the 1-, 2-, and 3-sigma intervals.}\label{fig:iraschi2}
\end{figure}
\begin{figure}
\centering
\includegraphics[width=.5\textwidth]{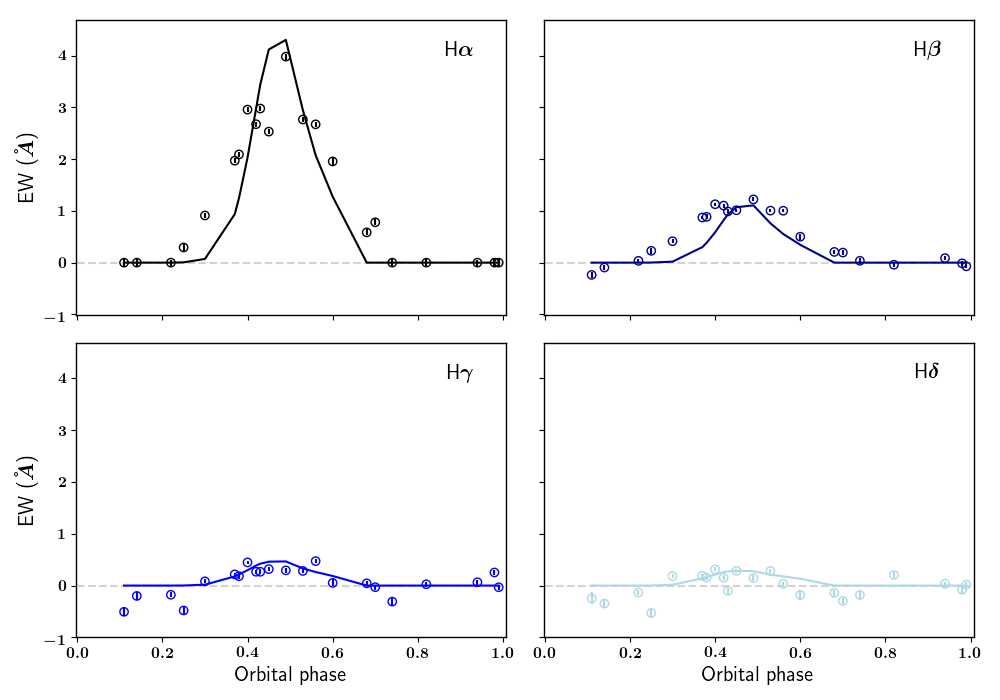}
\caption{As for Fig.~\ref{fig:bd_ewfit}, but for \iras.}\label{fig:iras_ewfit}
\end{figure}

The 2D $\chi^2$-distribution for the fitting is shown in Fig.~\ref{fig:iraschi2} and the associated EW of the model is shown in Fig.~\ref{fig:iras_ewfit}. We calculate the marginalised probability distributions for $n_j$ and $T_j$, given by Eqs.~(\ref{eq:prob1d_rho}) and (\ref{eq:prob1d_T}), from which we can determine the mean and standard deviations. This gives a jet density of $n_j = 1.0^{+0.5}_{-0.4}\,\times\,10^{16}\,\text{m}^{-3}$ and jet temperature of $T_j = 5330\pm180\,$K.

\section{Discussion}\label{sec:discussion}

By fitting the spatio-kinematic structure of the jet and estimating its density structure, we have gathered crucial information about the jet. We can now estimate how much mass is being ejected by the jet, which is essential for understanding the mass accretion onto the companion and determining the source feeding this accretion.  
\subsection{Jet mass-loss rate}\label{ssec:massloss}
The velocity and density structure of the jets, calculated by fitting the models, is used to estimate the mass ejection rate. The mass ejection rate of the jet for both systems is estimated by calculating how much mass passes through the jet at a height of $1\,$AU from the launch point. 

In the case of \bd, the velocity and density profiles are determined by the X-wind configuration (see Sect.~\ref{ssec:spatioIRAS}). We calculate the density at a height of $z=1\,$AU using Eq.~(\ref{eq:v_xwind}). The mass ejection rate can be found by the integral
\begin{equation}
    \dot{M}_\text{jet} = \int^R_0 \, \rho(r)\cdot v(r)\cdot 2\pi r \, \dd r, \label{eq:mdotcalc}
\end{equation}
with $r = 1\,\text{AU}\cdot\tan\theta$. The velocity at each location in the jet is defined by Eq.~(\ref{eq:v_xwind}). In this way we find a mass ejection rate of  $\dot{M}_\text{jet} = 7^{+3}_{-2}\times 10^{-7}\,\myr$ for \bd.

For \iras, the data is best fitted by a disk wind model (see Sect.~\ref{ssec:spatioBD}), whose  velocity profile is described by Eqs.~(\ref{eq:diskwind_in}) \& (\ref{eq:diskwind_out}) for the inner and outer jet regions, respectively. From Eq.~(\ref{eq:mdotcalc}), we find a mass ejection rate of $\dot{M}_\text{jet} = 2.0^{+2}_{-0.7}\times 10^{-6}\,\myr$. We have listed these values, as lower and upper limits, in Table~\ref{tab:mrates}.

\subsection{The ejection efficiency}\label{ssec:accr_eject}

By assuming an ejection efficiency $\dot{M}_\text{jet}/\dot{M}_\text{acc}$, we can link the jet mass-loss rate ($\dot{M}_\text{jet}$) to the accretion rate ($\dot{M}_\text{acc}$), and hence obtain a range of possible accretion rates onto the circum-companion disk. By doing so, we can assess if the mass transfer from either the post-AGB or the circumbinary disk, or both can contribute enough mass to the circum-companion disk in order to sustain the observed jet mass-loss rates. 

Ejection efficiency has not been determined for post-AGB binary systems, but the same theory, i.e., magneto centrifugal driving, applies to YSOs, which have been studied extensively \citep[and references therein]{ferreira07}. Moreover, disks in YSOs are comparable in size to those in our post-AGB binary systems \citep{hillen17} and their mass ejection rates are similar to the rates estimated for jets in post-AGB systems ($10^{-8} - 10^{-4}\,\myr$; \citealt{calvet98, ferreira07}). Current estimates of ejection efficiencies for T Tauri stars are in the range $\dot{M}_\text{jet}/\dot{M}_\text{acc} \sim 0.01-0.1$ \citep{cabrit07, cabrit09, nisini18}. This said, the spread in these values is large and some studies have even found ratios higher than $0.3$ \citep{calvet98, ferreira06, nisini18}. 

In this work, we adopt a wide range of ejection efficiencies for both of our post-AGB binary objects, according to the typical ranges found through observations of YSOs: $0.01 < \dot{M}_\text{jet}/\dot{M}_\text{accr} < 0.3$. Under these assumptions, and by using the jet mass-loss rates from Sect.~\ref{ssec:massloss}, the accretion rates onto the two companions are (see Table~\ref{tab:mrates}):
\begin{eqnarray}
\left\{
    \begin{array}{ll}
         1.7\times10^{-6}\,\myr < \,\dot{M}_\text{acc,BD} &< 1\times10^{-4}\,\myr \\
         5\times10^{-6}\,\myr < \,\dot{M}_\text{acc,IRAS} &< 4\times10^{-4}\,\myr. \label{eq:masslossboth}
    \end{array}
\right.
\end{eqnarray}

\noindent Next, we use these results to look at possible sources feeding the accretion.

\subsection{Sources of accretion onto the companion}\label{ssec:reaccretion}
\begin{table*}[h!]
\begin{center}
\caption{Derived accretion and mass-loss rates in the two binary systems.}
\label{tab:mrates}
\scalebox{0.85}{\begin{tabular}{l cccc}\\
\hline
\hline 
Parameter & \multicolumn{2}{c}{\bd} & \multicolumn{2}{c}{\iras} \\
 & lower limit  & upper limit & lower limit  & upper limit  \\
\hline
Jet mass-loss rate ($\myr$) &  $5\times 10^{-7}\,\myr$  & $1\times 10^{-6}\,\myr$  &  $1.3\times 10^{-6}\,\myr$ &  $4\times 10^{-6}\,\myr$ \\
Mass accretion rate onto the companion($\myr$) &  $1.7\times10^{-6}\,\myr$ & $1\times10^{-4}\,\myr$  & $4\times10^{-6}\,\myr$ &  $4\times10^{-4}\,\myr$ \\
mass-loss rate from the post-AGB star ($\myr$) &   & $<3.5\times 10^{-7}\,\myr$  &   & $<1.7\times 10^{-7}\,\myr$  \\
mass-loss rate from the circumbinary disk ($\myr$) &  $3\times10^{-8}\,\myr$  &  $6\times10^{-6}\,\myr$  &  $5\times10^{-8}\,\myr$  &  $9\times10^{-6}\,\myr$\\
\hline
\end{tabular}}
\end{center}
\tablefoot{The tabulated orbital parameters are: orbital period $P$, time of periastron $T_0$, eccentricity $e$, argument of periastron $\omega$, systemic velocity $\gamma$, radial velocity of the primary $K_\mathrm{1}$, projected semi-major axis of the primary $a_\mathrm{1}\sin i$, and mass function $f(\mathrm{m})$.}
\end{table*}
There are two possible sources of mass transfer onto the companion. The first is the post-AGB primary itself, moving mass via the first Lagrange point L1 and creating a circum-companion accretion disk. The second possibility is re-accretion of gas from the circumbinary disks \citep{vanwinckel03, deruyter06,dermine13}.

\subsubsection{Scenario 1: Mass transfer from the post-AGB star to the companion}

We first assume that the accretion onto the companion is due to mass transfer from the primary via L1. To estimate the mass transfer rate by the post-AGB star, we follow the prescription in \cite{ritter88}:
\begin{equation}
    \dot{M}_\text{1} = \frac{2\,\pi}{\sqrt{e}} \left( \frac{k_\text{B}}{m_\text{H}\,\mu_{1,\text{ph}}}T_1 \right)^{3/2} \frac{R_1^3}{GM_1} \, \rho_{1,\text{ph}} \, F(q), \label{eq:rlof}
\end{equation}
with $e$ Euler's number, $k_\text{B}$ the Boltzmann constant, $G$ the gravitational constant, $m_\text{H}$ the hydrogen mass, and  $\mu_{1,\text{ph}}$, $T_1$, $R_1$, $M_1$, and $\rho_{1,\text{ph}}$ the mean molecular weight, the temperature, the radius, and the mass, of the primary star, respectively.  
$F(q)$ is defined as 
\begin{equation}
    F(q) = \Bigg[ \Big(g(q)-(1+q)\Big)\,g(q) \Bigg]^{-1/2} \left(\frac{R_\text{1, RL}}{a}\right)^{-3},\label{eq:fq}
\end{equation}
with $q=M_2/M_1$ the mass ratio, $R_\text{1,RL}$ the Roche lobe radius of the post-AGB star, and $a$ the binary separation. In Equation~\ref{eq:fq} $g(q)$ is defined as
\begin{equation}
    g(q) = \frac{q}{x^3} + (1-x)^{-3},\label{eq:gq}
\end{equation}
with x the distance between the mass centre of the post-AGB star and L1 in terms of the the binary separation ($a$). 

We assume a neutral cosmic mixture which implies a mean molecular weight of $\mu_{1,\text{ph}}=0.8$. We note that this prescription is based on Roche lobe overflow for a star filling its Roche lobe, transferring mass to the companion. In this case, the radius of the star $R_1$ is equal to the Roche radius $R_\text{1,RL}$. However, from our results in in the spatio-kinematic modelling, the post-AGB stars in these two systems do not fill their Roche lobes, as is shown in the geometrical representation of the systems in Figs.~\ref{fig:bd_geometry} and \ref{fig:iras_geometry}. The observations also support this result, since a star that would fill at least $80\%$ of its Roche lobe, would show ellipsoidal variations in its light curve \citep{wislon76}. The light curves of \bd\, and \iras\, do not show these variations \citep{bollen19}. Hence, we extrapolate Eq.~(\ref{eq:rlof}) by using the radius of the primary $R_1$, instead of the Roche radius $R_\text{1,RL}$.

Additionally, since we find that the post-AGB star does not fill its Roche lobe, the mass transfer would occur via a mechanism less efficient and less strong than RLOF. A few other possibilities being wind-RLOF and Bondi-Hoyle-Lyttleton (BHL) accretion. In the case of wind-RLOF, the stellar wind will be focused to the orbital plane and most of the mass will be lost through the L1-point, towards the secondary \citep{mohamed07}. The mass-transfer efficiency of wind-RLOF would vary between a few percent and can reach up to $50\%$ \citep{devalborro09, abate13}. In the case of BHL accretion, the accretion efficiency will be significantly lower at about $1-10\%$ \citep{abate13, mohamed12}. Hence, the upper limit for mass transfer through wind-RLOF would be lower than that of RLOF. The upper limit for BHL accretion would be several orders of magnitude lower. Hence, by equating the mass-transfer from the post-AGB star using Eq.~(\ref{eq:rlof}), we can get a good estimation for an upper limit of mass-transfer from the post-AGB star to the companion.

To determine the photospheric density, we use the MESA stellar evolution code \citep[MESA][]{Paxton2011, Paxton2013, Paxton2015, Paxton2018, Paxton2019} to calculate the evolution of a post-AGB star with the correct mass. We subsequently use the photospheric density from the MESA output at the time-step when the post-AGB star has the same size as our star. This gives us a value of $10^{-10}\,\text{g\,cm}^{-2}$ for the photospheric density of the star ($\rho_{1,\text{ph}}$). The mass of the post-AGB star $M_1$ is set to $0.6\,M_\odot$, which is a typical value for these objects. The mass of the companion $M_2$ is determined from the mass function $f(M_1)$ and the inclination of the binary system found in the spatio-kinematic model fitting:
\begin{equation}
    f(M_1) = \frac{M_2^3\,\sin^3 i}{(M_1+M_2)^2}.\label{eq:fm}
\end{equation}
This gives a mass of $1.07\,M_\odot$ for the companion star of \bd, resulting in a mass ratio of $q=1.79$. We find the Roche radius of the post-AGB star using the formula by \cite{eggleton83}:
\begin{equation}
    R_\text{RL,1} = a\,\cdot\,\frac{0.49\,q^{-2/3}}{0.6\,q^{-2/3} + \ln\,(1+q^{-1/3})}.\label{eq:rocher}
\end{equation}
Using these values for Eq.~(\ref{eq:rlof}), we find the mass-transfer rate from the post-AGB star to the companion in \bd\, to be $\dot{M}_\text{1}=3.5\times 10^{-7}\,\myr$. This value is less than the lower limit of the jet mass-loss rate in \bd\, (see Table~\ref{tab:mrates}). Moreover, in Sect.~\ref{ssec:massloss}, we found the mass accretion rate to be in the range of $1.7\times10^{-6}\,\myr <  \dot{M}_\text{accr} < 1\times10^{-4}\,\myr$, which is about five times the theoretical value for the upper-limit of mass-transfer from the post-AGB star to the companion. Hence, it is unlikely that the post-AGB star can contribute enough mass to the circum-companion disk to sustain the observed jet outflow.

We conduct a similar analysis for \iras\, and we come to a similar conclusion. The mass of the companion is $M_2=0.46\,M_\odot$ and the mass ratio would be $q=0.77$. We use the same values for $\mu_{1,\text{ph}}$ and $\rho_{1,\text{ph}}$, giving us a mass-transfer rate of $\dot{M}_\text{RLOF}=1.7\times 10^{-7}\,\myr$. Hence, this upper limit for mass-transfer from the post-AGB star for \iras\, is also smaller than the lower limit for the jet mass-loss rate of \iras\, and thus too low to match a measured mass accretion rate in the range $4\times10^{-6}\,\myr <  \dot{M}_\text{accr} < 4\times10^{-4}\,\myr$.\\

We conclude that it is unlikely that the mass transfer from the post-AGB star alone is responsible for feeding the accretion disk around the companion. 

\subsubsection{Scenario 2: Re-accretion from the circumbinary disk}
Here, we consider the possibility of mass accretion from the circumbinary disk onto the central binary system. In order to give an estimate of re-accretion by the circumbinary disk of \bd\, and \iras, we use the mass-loss equation by \cite{rafikov16}, that defines the mass loss by the disk to the central binary as a function of time:
\begin{equation}
    \dot{M}_\text{disk}(t) = \frac{M_{0,\text{disk}}}{t_0} \left( 1 + \frac{t}{2t_0} \right)^{-3/2},\label{eq:masslossdisk}
\end{equation}
where $M_{0,\text{disk}}$ the initial disk mass. These circumbinary disks have average disk masses of $M_{0,\text{disk}}=10^{-2}\,M_\odot$ \citep{gielen07, bujarrabal13a, bujarrabal18, hillen17, kluska18}. \cite{bujarrabal13a} and \cite{bujarrabal18} derived disk masses of circumbinary disks of post-AGB binary systems ranging from $6\times10^{-4}$ to $5\times10^{-2}\, M_\odot$. We will use this range to estimate the mass-loss rate from the disk.

The initial viscous time of the disk $t_0$ is defined by \cite{rafikov16} as
\begin{equation}
    t_0 = \frac{4}{3}\frac{\mu}{k_\text{B}}\frac{a_b}{\alpha} \left[ \frac{4\pi\sigma(GM_b)^2}{\zeta L_1} \right]^{1/4} \left( \frac{\eta}{I_L}, \right)^2
\end{equation}
where $\mu$ is the mean molecular weight, $a_b$ is the binary separation, $\alpha$ is the viscosity parameter, $\sigma$ is Stefan-Boltzmann constant, $L_1$ is the luminosity of the post-AGB star, and  $\zeta$ is a constant factor that accounts for the starlight that is intercepted by the disk surface at a grazing incidence angle. $\eta$ is the ratio of angular momentum of the disk compared to that of the central binary and $I_L$ characterises the spatial distribution of the angular momentum in the disk. We fix several values at the same values as \cite{rafikov16} and \cite{oomen19}: $\mu=2m_p$, $\alpha=0.01$, $\zeta=0.1$, and $I_L=1$, with $m_p$ the mass of a proton. The luminosity $L_1$ of \bd\, and \iras\, are $2100^{+1500}_{-800}\,$L$_\odot$ and $2100^{+500}_{-400}\,$L$_\odot$, respectively \citep{oomen19}. The angular momentum of the circumbinary disk is typically of the order of the angular momentum of the central binary system \citep{bujarrabal18, izzard18}. We will set a range of $\eta$ between $1.4-2$, where a value of $1.4$ is appropriate for a disk with the bulk of its mass located at the inner disk rim.

Using Eq.~(\ref{eq:masslossdisk}) and assuming $t=0$, we can calculate a range of possible mass-loss rates by the disk. We find a range of $3\times10^{-8}\,\myr <  \dot{M}_\text{disk} < 6\times10^{-6}\,\myr$ for \bd, while in the case of \iras, we find that the re-accretion rate from the circumbinary disk is in the range of $5\times10^{-8}\,\myr <  \dot{M}_\text{disk} < 9\times10^{-6}\,\myr$ (Table~\ref{tab:mrates}).  When the disk matter falls onto the central binary, it will be accreted by both the post-AGB star and the companion. Hence, the mass lost by the circumbinary disk should be twice the mass accreted by the circum-companion accretion disk. If we compare the mass accretion rate for \bd\, and \iras\, with the estimated mass-loss rate by the circumbinary disk, it shows that re-accretion from the circumbinary disk is a plausible mechanism for the formation of the jet. 

We note that only the higher estimates for mass accretion rates from the circumbinary disk can explain our observationally-derived rates. Hence, this would imply that for these two systems the disk masses are at the high end of the range ($M_{0,\text{disk}} > 10^{-2}\,M_\odot$) and that we are observing the early stages of the re-accretion by the circumbinary disk. Nevertheless, our findings are in good agreement with \cite{oomen19}, who estimated that accretion rates should be higher than $3\times10^{-7}\,\myr$ and that disk masses should be higher than $\sim 10^{-2}\,M_\odot$.\\ 


\section{Summary and conclusion}\label{sec:conclusions}

In this paper, we aimed to determine mass-transfer rates of jet-creating post-AGB binaries. We fully exploited the time-series of high-resolution optical spectra from these binary systems. We presented a new radiative transfer model for these jets and applied this model to reproduce the Balmer lines of two well-sampled post-AGB binary systems, i.e. \bd\, and \iras. With this model, we were able to study the mass-loss rate of the jet and mass accretion rate onto the companion, and constrain the source of the accretion in these systems: the post-AGB star or the circumbinary disk. Additionally, we expanded the spatio-kinematic model from \cite{bollen19}. Our main conclusions can be summarised as follows:
\begin{enumerate}
    \item  We successfully reproduced the observed absorption feature in the H$\alpha$ line profiles of our test sources with our improved spatio-kinematic model of the jet. By doing so, we obtained the kinematics and three-dimensional morphology of the jet. The implementation of the jet tilt in the model reproduced the observed lag of the absorption feature in the Balmer lines. This tilt is significant for both objects, with values of $15\,\degr$ and $6\,\degr$ for \bd\, and \iras, respectively. Likewise, the new jet cavity in the model improves the jet representation, as was suggested by \cite{bollen19}.
    
    \item We showed that we can acquire a three-dimensional jet morphology by modelling the amount of absorption in the H$\alpha$ lines from our spatio-kinematic model of the jet. By combining the results of the spatio-kinematic and radiative transfer modelling, we found the crucial parameters to calculate jet mass-loss rates, i.e. jet velocity and geometry from the spatio-kinematic model and jet density structure from the radiative transfer model.
    
    \item We computed the mass-loss rate of the jet by combining the results of our spatio-kinematic model and radiative transfer model. The computed mass-loss rates for the jets in \bd\, and \iras\, range between $(5-10)\times 10^{-7}\,\myr$ and $(1.3-4)\times 10^{-6}\,\myr$, respectively, as tabulated in Table~\ref{tab:mrates}. 
     These mass ejection rates are comparable to the mass ejection rates for the jets in planetary nebulae and pre-planetary nebulae \citep{tocknell14, tafoya19}. \cite{tocknell14} found the mass ejection rates to be $1 - 3 \times 10^{-7}\,\myr$ and $8.8 \times 10^{-7}\,\myr$ for the Necklace and NGC~6778, respectively. 
     
     These mass ejection rates imply correspondingly high mass accretion rates onto the companion that range between $1.7\times 10^{-6}\,\myr$ and $1\times 10^{-4}\,\myr$ for \bd\, and $4\times 10^{-6}\,\myr$ and $4\times 10^{-4}\,\myr$ for \iras.
    
    \item By determining the jet mass-loss rate we added an additional constraint on the nature of the accretion onto these systems. While the uncertainties are high, the circumbinary disk is the preferred source of accretion feeding the jet rather than the post-AGB star: the accretion rates from the post-AGB stars are too low to justify the observed jet mass-loss rates. We note, however, that the simultaneous accretion from both the circumbinary disk and the post-AGB star cannot be ruled out. Re-accretion from the circumbinary disk also naturally explains the abundance pattern of the post-AGB star and is in agreement with the study by \cite{oomen19}, who showed that high re-accretion rates ($> 3\times 10^{-7}\,\myr$) are needed in order to reproduce the observed depletion patterns of post-AGB stars. These high re-accretion rates from the circumbinary disk can prolong the lifetime of the post-AGB star and thus have an important impact on the evolution of these objects, provided that the disk can sustain the mass-loss.

\end{enumerate}
In our future studies, we will perform a comprehensive analysis of the whole diverse sample of jet-creating post-AGB binary systems, by using both the spatio-kinematic and radiative transfer models. The observational properties of these binaries and their jets are in-homogeneous. Hence, by analysing the whole sample, we aim to  obtain strong constraints on the source of the accretion and identify correlations between mass-accretion, depletion patterns, and the orbital properties of post-AGB binaries.

\begin{acknowledgements}
This work was performed on the OzSTAR national facility at Swinburne University of Technology. OzSTAR is funded by Swinburne University of Technology and the National Collaborative Research Infrastructure Strategy (NCRIS). DK  acknowledges  the  support  of  the  Australian  Research Council  (ARC)  Discovery  Early  Career  Research  Award (DECRA) grant (95213534). HVW acknowledges support from the Research Council of the KU Leuven under grant number C14/17/082.  The observations presented in this study are obtained with the HERMES spectrograph on the Mercator Telescope, which is supported by the Research Foundation - Flanders (FWO), Belgium, the Research Council of KU Leuven, Belgium, the Fonds National de la Recherche Scientifique (F.R.S.-FNRS), Belgium, the Royal Observatory of Belgium, the Observatoire de Gen\`eve, Switzerland and the Th\"uringer Landessternwarte Tautenburg, Germany.
\end{acknowledgements}

\bibliographystyle{aa}
\bibliography{allreferences.bib}

\begin{appendix}

\section{The absorption feature in the H$\alpha$ line.}

\begin{figure}[h!]
\centering
  \includegraphics[width =.5\textwidth]{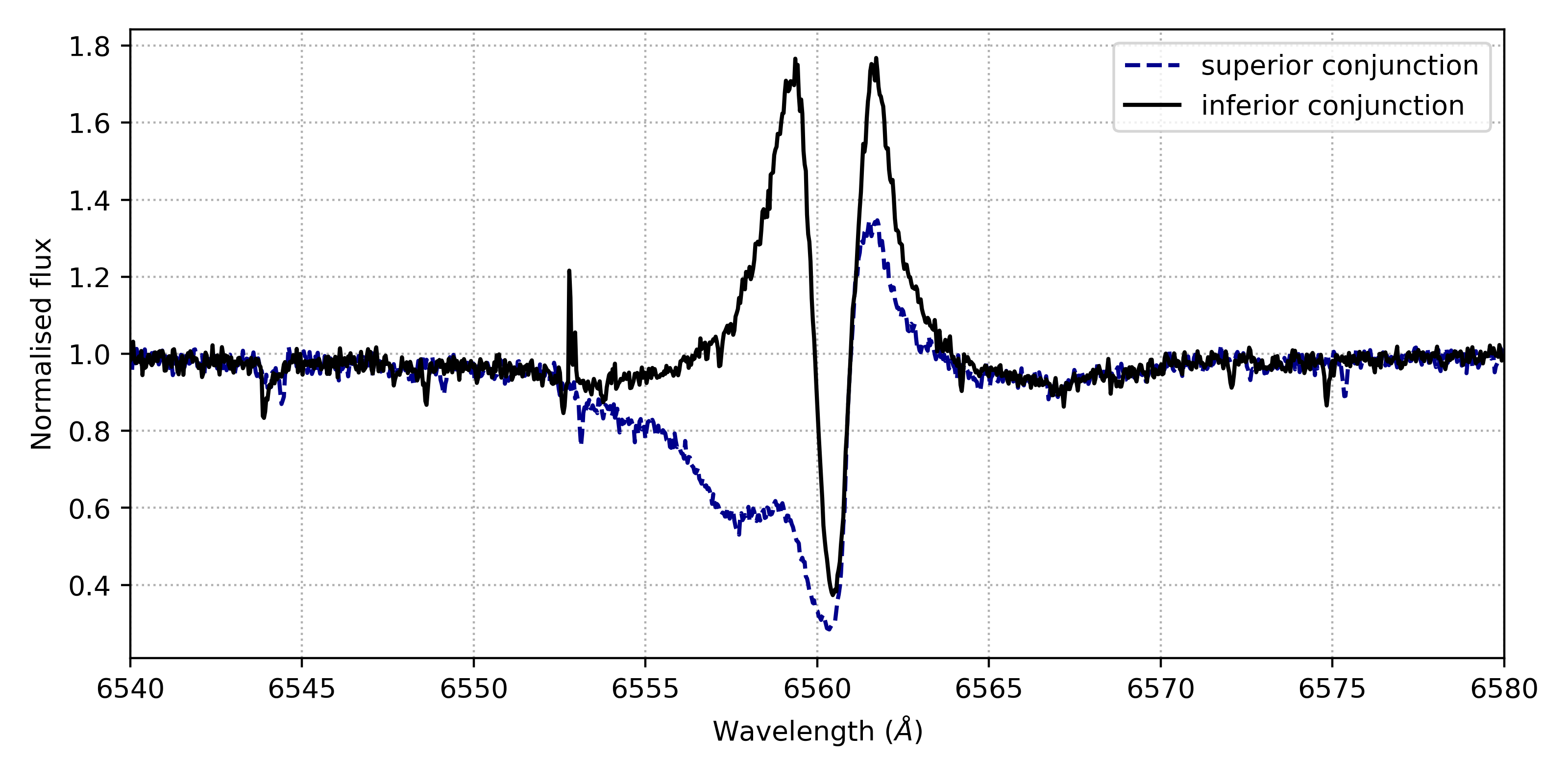}
\caption{H$\alpha$-line of \bd\, at two different phases in the orbital period. The H$\alpha$ line displays a double-peaked emission feature with a central absorption feature during inferior conjunction (the black solid line), when the post-AGB star is between the jet and the observer. During superior conjunction, when the jet is between the post-AGB star and the observer, we observe a blue-shifted absorption feature in the H$\alpha$ line (the blue dotted line).}\label{fig:supinf}
\end{figure}

\section{Orbital parameters of \bd\, and \iras.}\label{ap:orbpar}
\begin{table}[h!]
\begin{center}
\caption{Spectroscopic orbital solutions of the primary component of \bd\, and \iras\, \citep{oomen18}.}
\label{tab:orbpar}
\begin{tabular}{l cc}\\
\hline
\hline 
Parameter & \bd & \iras \\
\hline
$P$ (d)                                 &  $140.82\pm0.02$      &   $126.97\pm0.08$\\
$T_0$ (BJD)                             &  $2455233.9\pm1.3$    &   $2454997.7\pm1.0$\\
$e$                                     &  $0.085\pm0.005$      &   $0.13\pm0.03$\\
$\omega$ ($^{\circ}$)                   &  $95.8\pm3.3$         &   $66.0\pm4.4$\\
$\gamma$ (km s$^{-1}$)                  &  $-98.13\pm0.8$       &   $31.7\pm0.3$  \\
$K_1$ (km s$^{-1}$)         &  $23.8\pm0.1$         &   $18.0\pm0.6$\\
$a_1\sin i$ $(\mathrm{AU})$ &  $0.3074\pm0.0014$    &   $0.209\pm0.008$\\
$f(\mathrm{m})$ $(M_\odot)$             &  $0.195\pm0.003$      &   $0.075\pm0.008$\\
\hline
\end{tabular}
\end{center}
\tablefoot{The tabulated orbital parameters are: orbital period $P$, time of periastron $T_0$, eccentricity $e$, argument of periastron $\omega$, systemic velocity $\gamma$, radial velocity of the primary $K_\mathrm{1}$, projected semi-major axis of the primary $a_\mathrm{1}\sin i$, and mass function $f(\mathrm{m})$.}
\end{table}

\section{Balmer lines of \bd\, and \iras.}\label{ap:balmerlines}
\begin{figure*}
\centering
\includegraphics[width=1\textwidth]{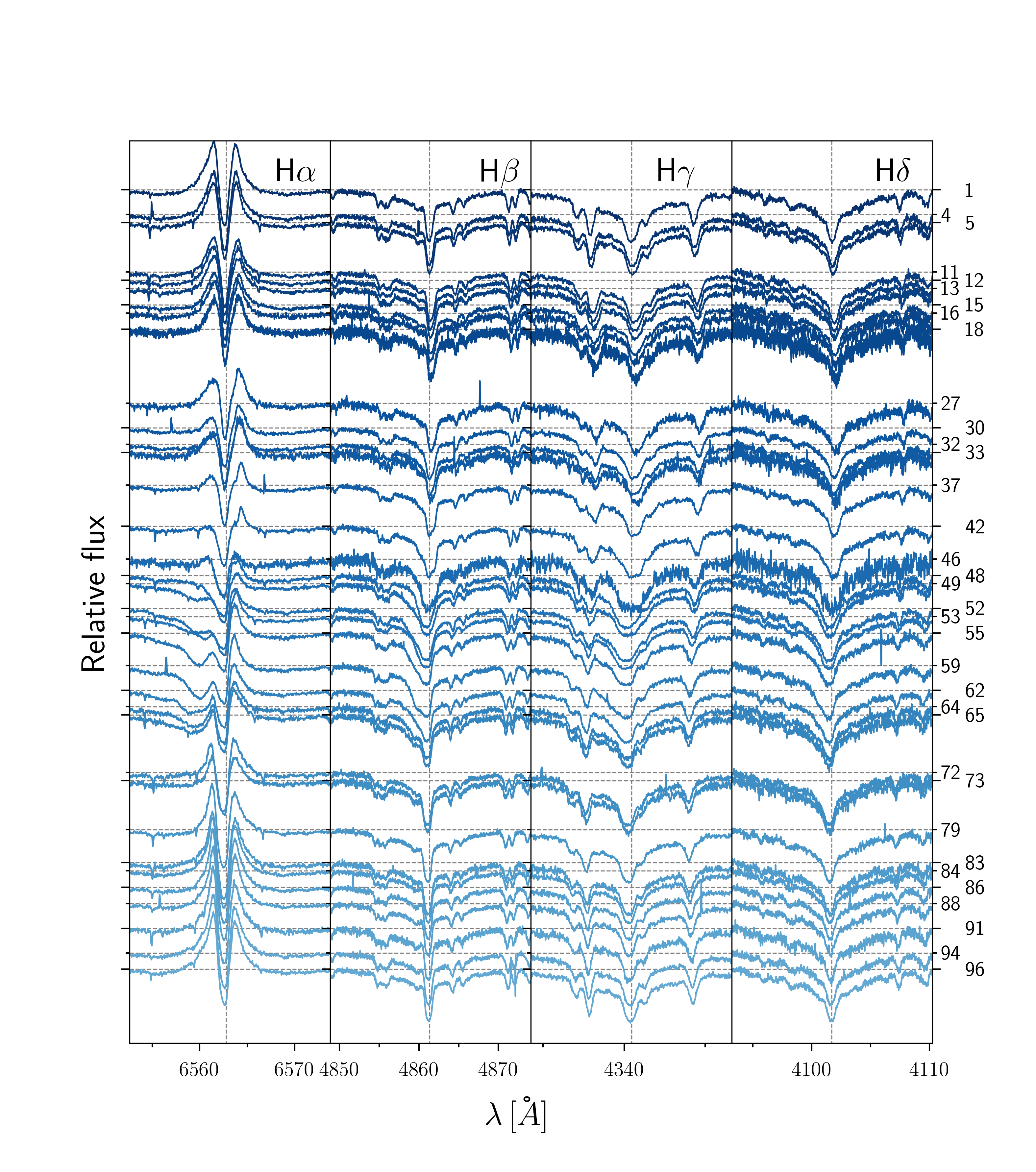}
\caption{Balmer lines of \bd\, as a function of wavelength. The spectra are given in arbitrary units and offset according to their orbital phase. Numbers on the right vertical axis indicate the orbital phase of the spectra (from $0-100$). The dashed vertical lines represent the centre of each Balmer line.}\label{fig:bd_balmer}
\end{figure*}

\begin{figure*}
\centering
\includegraphics[width=1\textwidth]{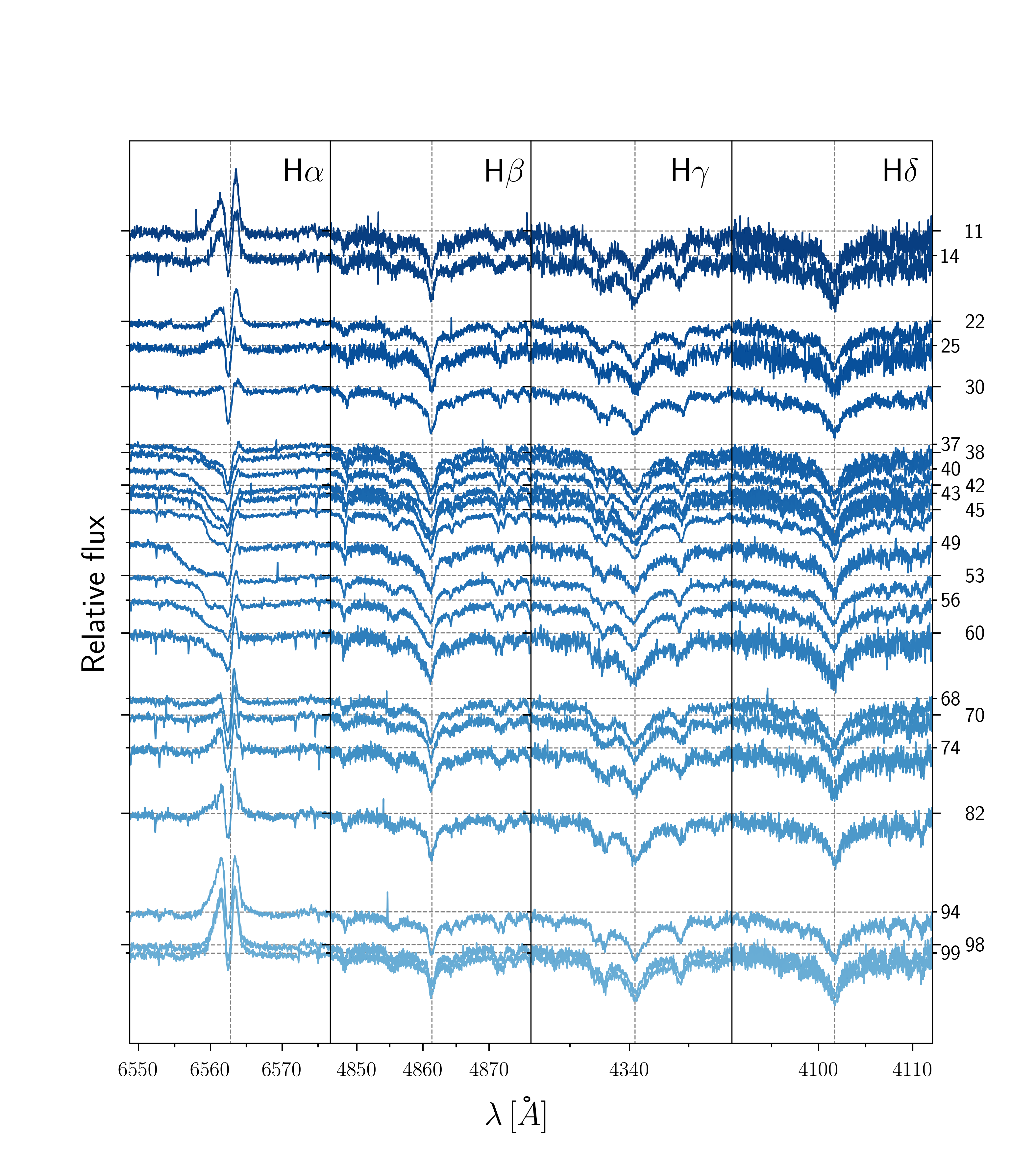}
\caption{Similar to Fig.~\ref{fig:bd_balmer}, but for \iras.}\label{fig:iras_balmer}
\end{figure*}


\end{appendix}

\end{document}